\documentclass[aps,prb,twocolumn,showpacs,superscriptaddress,floatfix]{revtex4}
\pdfoutput=1

\usepackage{graphicx}
\usepackage{amsmath,amssymb}

\usepackage{ulem}
\usepackage{color}
\usepackage{amssymb}

\def\diff{\mathrm d}
\def\mathi{\mathrm i}

\newcommand{\opcdag}{\ensuremath{c^\dagger}}
\newcommand{\opc}{\ensuremath{c}}

\newcommand{\boldr}{\boldsymbol{r}}

\newcommand{\Ud}{U_d}

\begin{document}
\title{Accuracy of downfolding based on the constrained random phase approximation}
\author{Hiroshi Shinaoka} 
\affiliation{Theoretische Physik, ETH Z\"{u}rich, 8093 Z\"{u}rich, Switzerland}
\affiliation{Department of Physics, University of Fribourg, 1700 Fribourg, Switzerland}
\author{Matthias Troyer}
\affiliation{Theoretische Physik, ETH Z\"{u}rich, 8093 Z\"{u}rich, Switzerland}
\author{Philipp Werner}
\affiliation{Department of Physics, University of Fribourg, 1700 Fribourg, Switzerland}
\date{\today}

\begin{abstract}
  We study the reliability of the constrained random phase approximation (cRPA) method for the calculation of low-energy effective Hamiltonians by considering multi-orbital lattice models with one strongly correlated ``target" band and two weakly correlated ``screening" bands.
  The full multi-orbital system and the effective model are solved within dynamical mean field theory (DMFT) in a consistent way.
  By comparing the quasi-particle weights for the correlated bands,
  we examine how accurately the effective model describes the low-energy properties of the multi-band system.
  We show that the violation of the Pauli principle in the cRPA method leads to overscreening effects when the inter-orbital interaction is small.
  This problem can be overcome by using a variant of the cRPA method which restores the Pauli principle. 
\end{abstract}


\maketitle


\section{Introduction}
Strongly correlated electron systems attract much attention because they exhibit remarkable many-body phenomena.
Establishing a first-principles theoretical framework for describing the electronic properties of this class of materials is a great challenge.  
Methods based on density functional theory (DFT)~\cite{Hohenberg:1964zz,Kohn:1965ui} have been successfully used to understand and predict the properties of weakly correlated materials such as elemental metals.
Although the DFT formalism is exact in principle,
the density functionals used in calculations are approximate, e.g., based on the local density approximation (LDA),~\cite{Kohn:1965ui} because the exact form of the functional is unknown.
The result is a static mean-field description of the electronic structure.
Applying this approach to strongly correlated materials misses fundamental aspects, such as quantum fluctuations and Mott physics. 

On the other hand, a variety of sophisticated numerical methods have been developed to treat effective models of strongly correlated electrons in lattice systems such as the Hubbard model.
Examples include quantum Monte Carlo methods,~\cite{Gull:2011jda} dynamical mean-field theory (DMFT),~\cite{Georges:1996un}
the variational Monte Carlo method,~\cite{Ceperley:1977zz}
density matrix renormalization group,~\cite{ck:2011gl} and tensor network methods.
These methods take into account correlation effects beyond the static mean-field level.
However, they cannot be directly applied to real materials, 
which are typically characterized by a complex and hierarchical electronic structure.
In most transition metal oxides, there are only a few correlated bands near the Fermi level, which are typically of $d$ character, and in the simplest situation 
these bands are well separated in energy from the higher- and lower-lying bands (which we will collectively denote as ``high-energy bands"). 
Although high-energy bands are usually less correlated, 
they can substantially affect the low-energy electrons through the screening of the Coulomb interaction.
Thus, we cannot simply neglect the high-energy degrees of freedom in realistic calculations. 
A similar structure is found also in lanthanide or actinide oxides and organic compounds.

In recent years much effort has been devoted to establishing reliable first-principles methods for strongly correlated materials which exploit this hierarchical structure.~\cite{Imada:2010ep}
The strategy is to construct an effective low-energy lattice model, which contains only a few degrees of freedom,
by eliminating the high-energy degrees of freedom in a systematic manner. 
In practice, we compute effective Coulomb interactions in the low-energy model by taking into account the screening effects by the high-energy bands using a first-principles calculation based on DFT.
Then, this strongly correlated effective model is solved accurately by quantum Monte Carlo methods or DMFT. 
The procedure which leads to the low-energy effective model is called downfolding.
One widely used method for computing screening effects is the constrained random approximation (cRPA) method.~\cite{Aryasetiawan:2004hd}
It has been applied to a variety of transition metal oxides~\cite{Nakamura:2010jr,Miyake:2010eo,Nakamura:2008boa,Misawa:2012jo} and organic compounds~\cite{Nakamura:2009dm,Nakamura:2012co,Shinaoka:2012kl,Koretsune:2014id} to investigate metal-insulator transitions, magnetism, and superconductivity.

While the logic behind the cRPA method is compelling,
it is at the present stage a recipe, whose accuracy and limitations have not been established.
To the best of our knowledge, no systematic effort has yet been made to clarify under which circumstances and to what extent cRPA is reliable. 
An obvious difficulty is that an accurate numerical solution of the original multi-band problem is in general not possible. 
For this reason we address the issue in a simple model context where the accuracy of the cRPA downfolding scheme can be tested systematically.
We consider multi-orbital Hubbard models in three dimensions and derive effective low-energy models by the cRPA downfolding scheme.
Then, we solve both the full model and the effective model using a DMFT or extended DMFT approximation. By comparing quantities such as mass enhancements, we can determine the parameter regions in which the effective model provides an accurate description of the low-energy properties of the original multi-band model.
It is generally expected that cPRA works best if the screening bands are at high energies.~\cite{Imada:2010ep}
However, this ideal situation is not realized in many relevant materials such as high-$T_c$ cuprates~\cite{Werner:2014tm} and correlated organic compounds.~\cite{Nakamura:2009dm,Nakamura:2012co}
In order to understand the limitations of the cRPA method,
we will focus in this study on models with a few screening bands which are close in energy to the target band.

The rest of this paper is organized as follows. 
In Sec.~II, we introduce the model used in this study.
In Sec.~III, we explain the cRPA downfolding procedure.
Section~IV describes the details of the DMFT calculations.
We discuss results of the downfolding and DMFT calculations in Sec. V.
Section~VI contains the conclusions and a brief outlook.

\section{Model}\label{sec:model}
\begin{figure}
 \centering\includegraphics[width=.45\textwidth,clip]{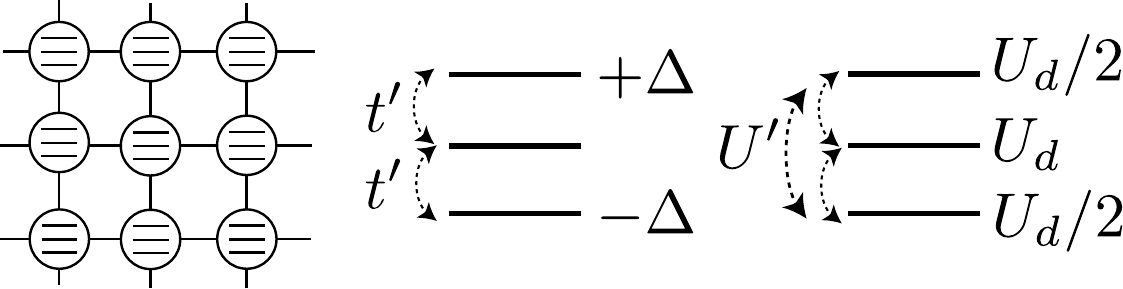}
 \caption{
 Three-orbital model on a cubic lattice.
 The three levels are split by orbital dependent on-site energies.
 We include an orbital-offdiagonal transfer $t^\prime$, but the highest and lowest orbitals are not connected by a hopping term.
 The on-site repulsion for the target orbital is denoted by $U_d$, while the highest and lowest orbitals are less correlated with an on-site repulsion of $U_d/2$.
 The inter-orbital repulsion $U^\prime$ acts between all pairs of orbitals.
 }
\label{fig:model}
\end{figure}

\begin{figure}[t]
 \centering\includegraphics[width=.5\textwidth,clip]{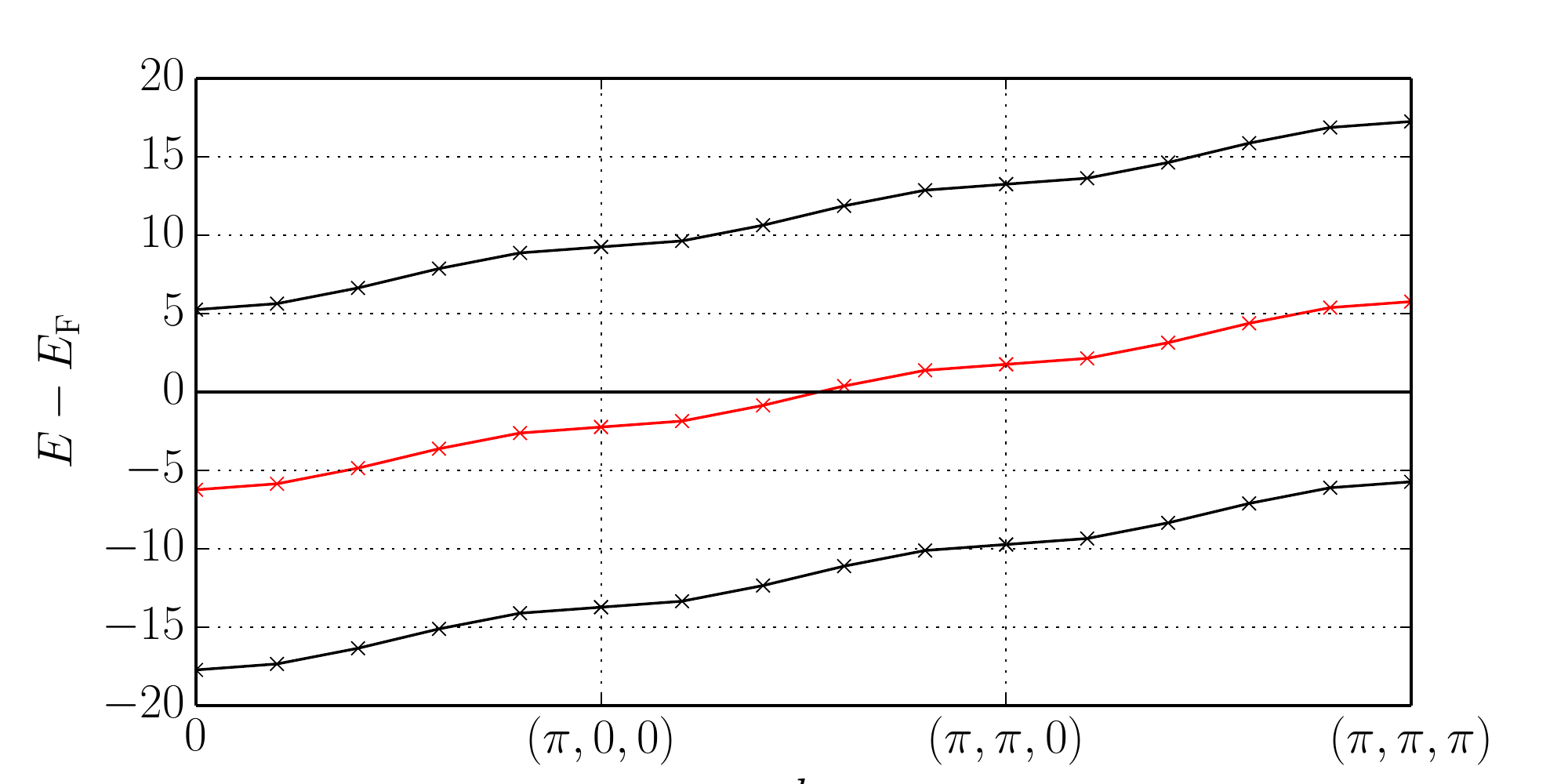}
 \caption{
 (Color online) 
 Non-interacting band structure of the three-orbital model for $\Delta=10$ and $t^\prime=4$.
 The half-filled target bands are shown in red.
 }
 \label{fig:band}
\end{figure}

To test the accuracy of downfolding,
we consider a three-orbital Hubbard model on a cubic lattice with orbital-diagonal transfer $t=1$ between nearest-neighbor sites.
Its Hamiltonian is given by
\begin{eqnarray}
  \mathcal{H} &=& - \sum_{\langle i,j\rangle}\sum_{\alpha\sigma} \hat{c}^\dagger_{i\alpha\sigma}\hat{c}_{j\alpha\sigma} + \sum_i \sum_\alpha (E_\alpha+E^\mathrm{dc}_\alpha-\mu) \hat{n}_{i\alpha} \nonumber \\
  &&-t^\prime \sum_i \sum_{\sigma}\sum_{\beta\neq 2} \left(\hat{c}^\dagger_{i2\sigma}\hat{c}_{i\beta\sigma} + \hat{c}^\dagger_{i\beta\sigma}\hat{c}_{i2\sigma} \right)\nonumber\\
  &&+\sum_{i\alpha} U_\alpha \hat{n}_{i\alpha\uparrow}\hat{n}_{i\alpha\downarrow} + \sum_i \sum_{\alpha<\beta} U^\prime \hat{n}_{i\alpha} \hat{n}_{i\beta},\label{eq:Ham}
\end{eqnarray}
where $i$ and $j$ are site indices, while $\alpha$ and $\beta$ are orbital indices.
We consider only density-density interactions.
The on-site repulsion $U_\alpha$ is taken to be $U_\alpha=U_d/2,~U_d,~U_d/2$ for $\alpha=1,2,3$, respectively ($U_d>0$) because screening bands are usually less correlated than target bands in real materials (see illustration in Fig.~\ref{fig:model}).
We also include inter-orbital interactions $U^\prime$.\
The chemical potential $\mu$ is adjusted in the DMFT self-consistent procedure such that the number of electrons is 3 (half filling).
The orbital potentials $E_\alpha$ are given by $-\Delta$, $0$, $\Delta$ for $\alpha=1,2,3$.
$\Delta>0$ produces gaps between the target- and screening-band manifolds.
We show the non-interacting band structure for $\Delta=10$ and $t^\prime=4$ in Fig.~\ref{fig:band}.
The half-filled target band is sandwiched between two high-energy bands.
Although the target band and the screening bands are separated by a direct gap at each $k$ point, the indirect gap is negative.
The Coulomb interaction breaks the particle-hole symmetry because it induces orbital-dependent mean fields.
To retrieve this symmetry in the limit of $t^\prime=0$,
we take $E^\mathrm{dc}_\alpha=U^\prime$, 0, $-U^\prime+U_d/2$ for $\alpha=1,2,3$.
For more details, we refer to Appendix~\ref{sec:band-shift}.
Indeed, the Hartree-Fock band structure remains almost symmetric in the parameter regime considered in this paper.
As we will explain later, the polarization function is computed using the Hartree-Fock band structure in the cRPA downfolding procedure.

\section{Constrained random phase approximation (cRPA)}
In this section, we describe the details of the cPRA formalism used in this study. 
In Sec.~\ref{sec:conv-cRPA}, we review the spin-independent formalism,
which is usually used for first-principle calculations.
We start from a real-space formalism, and derive the cRPA equation in a tight-binding form.
Section~\ref{sec:spin-dep-cRPA} describes the extension to a spin-dependent formalism,
where the spin dependence of the intra-orbital interactions is taken into account.

\subsection{Spin-independent formalism}\label{sec:conv-cRPA}
We start by considering the Hamiltonian 
\begin{eqnarray}
  \mathcal{H}&=& H^0 + V, \\
  H^0 &=& \sum_{n=1}^N h^0(r_n) = \sum_{n=1}^N \left[ -\frac{1}{2}\nabla^2_n + V_\mathrm{ext}(r_n)\right], \\
  V &=& \frac{1}{2}\sum_{i\neq j} v(r_i-r_j),\label{eq:ham-cont}
\end{eqnarray}
where $r_n$ is a combined index for the position and spin of an electron, i.e. $r_n\equiv (\boldsymbol{r}_n,\sigma_n)$, and assume that $v(r_i-r_j)$ is a spin-independent two-body Coulomb interaction.

In second quantization, this Hamiltonian reads
\begin{eqnarray}
  \mathcal{H} &=& \sum_{ij} t_{ij} c^\dagger_{i\sigma} c_{j\sigma} + \hat{V},
\end{eqnarray}
where $i$ and $j$ are indices of an orthonormal single-particle basis $\{\phi_{i\sigma}\}$, and $\sigma$ denotes the spin.
For convenience, we assume that $\phi_{i\sigma}$ has non-zero elements only in the spin sector $\sigma$.
The Coulomb interaction has the form
\begin{eqnarray}
  \hat{V} &=& \frac{1}{2}\sum_{ijkl} \sum_{\sigma_1\sigma_2\sigma_3\sigma_4} V_{ijkl}^{\sigma_1\sigma_2\sigma_3\sigma_4} c^\dagger_{i\sigma_1} c^\dagger_{j\sigma_2} c_{k\sigma_3} c_{l\sigma_4},
\label{V}
\end{eqnarray}
with
\begin{eqnarray}
  V_{ijkl}^{\sigma_1\sigma_2\sigma_3\sigma_4} &=& \int\diff r \diff r^\prime  \phi^*_{i\sigma_1}(r) \phi^*_{j\sigma_2}(r^\prime) v(\boldsymbol{r}-\boldsymbol{r}^\prime)\nonumber\\
  && \times \phi_{k\sigma_3}(r^\prime) \phi_{l\sigma_4}(r) \label{eq:Vijkl}
\end{eqnarray}
and $c_{i\sigma}$ and $c^\dagger_{i\sigma}$ the annihilation and creation operators for the single-particle basis, respectively.
In practice, we consider density-density terms.
For example, the on-site repulsion $U_\alpha$ and the inter-orbital interaction $U^\prime$ are represented by
\begin{eqnarray}
  U_\alpha &=& \int \mathrm{d} \boldsymbol{r} \mathrm{d} \boldsymbol{r}^\prime \phi_{i\alpha}^2(\boldsymbol{r}) V(\boldsymbol{r}, \boldsymbol{r}^\prime) \phi_{i\alpha}^2(\boldsymbol{r}^\prime),\\
  U^\prime &=& \int \mathrm{d} \boldsymbol{r} \mathrm{d} \boldsymbol{r}^\prime \phi_{i\alpha}^2(\boldsymbol{r}) V(\boldsymbol{r}, \boldsymbol{r}^\prime) \phi_{i\beta}^2(\boldsymbol{r}^\prime).
\end{eqnarray}
\begin{figure}
 \centering\includegraphics[width=.3\textwidth,clip]{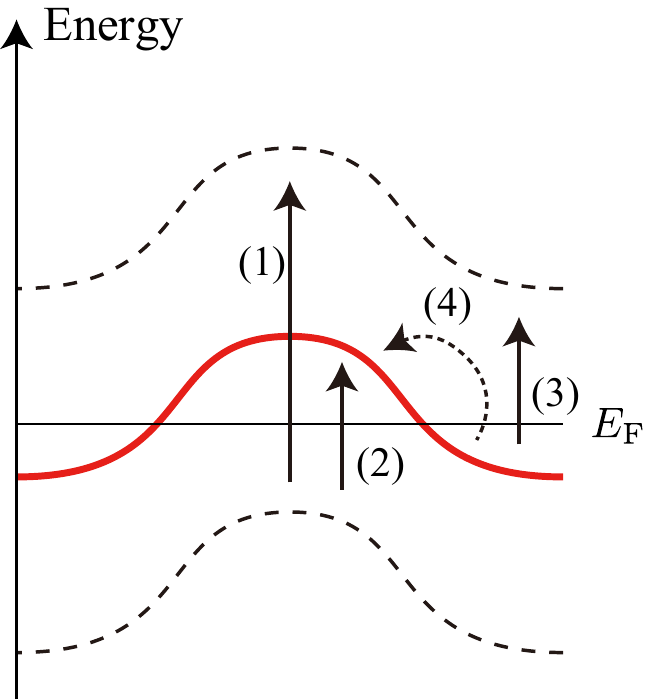}
 \caption{
 Schematic band structure.
 The solid line denotes the low-energy band in the target manifold of the low-energy model, while screening bands are denoted by broken lines.
 Solid and broken arrows show possible contributions to the polarization:
 excitations between 
 (1) occupied screening bands and unoccupied screening bands,
 (2) occupied screening bands and unoccupied target bands,
 (3) occupied target bands and unoccupied screening bands,
 (4) occupied target bands and unoccupied target bands.
 In the cRPA method, we exclude the contribution of (4) because this should be taken into account in solving the low-energy effective model.
 }
 \label{fig:cRPA}
\end{figure}

Now, let us consider a non-interacting band structure in which only a few bands are crossing the Fermi level and these low-energy target bands are sandwiched by high-energy screening bands.
Figure~\ref{fig:cRPA} illustrates a simple example, which has one target band and two screening bands.
In the downfolding, we derive an effective model for the target manifold by integrating out the high-energy screening bands.
The effective model has the form 
\begin{eqnarray}
  \mathcal{H} &=& \sum_{ij} \bar{t}_{ij} d^\dagger_{i\sigma} d_{j\sigma} + \hat{W},
\end{eqnarray}
where we introduced a new single-particle basis $\{\bar{\phi}_{i\sigma}\}$ and hopping parameters $\bar{t}_{ij}$ to describe the band structure in the target manifold. 
The annihilation and creation operators of $\{\bar{\phi}_{i\sigma}\}$ are given by $d$ and $d^\dagger$, respectively.
We usually take $\bar{\phi}$ to be localized in real space so that the effective interaction is as short-ranged as possible.
In first-principles calculations based on a plane-wave basis, the maximally localized Wannier functions~\cite{Souza:2001ba,Marzari:329970} are a common choice.

The screened interaction $w(r,r^\prime)$ is given by
\begin{eqnarray}
  w(r,r^\prime)&=& v(r,r^\prime) \label{eq:screening-rs}\\
  && + \int \mathrm{d} r_1 \mathrm{d} r_2 v(r,r_1) P(r_1,r_2) w(r_2,r^\prime),\nonumber
\end{eqnarray}
where the polarization $P$ is calculated within the bubble approximation (neglecting vertex corrections) as\cite{Aryasetiawan:2004hd}
\begin{eqnarray}
	&&P(r,r^\prime;\omega)= \sum_{n,m}^\prime f(\epsilon_n)(1-f(\epsilon_m))\times\nonumber \\
        &&\hspace{5mm}\left(
        \frac{A_{n,m}(r,r^\prime)}{\omega-(\epsilon_m-\epsilon_n)+\mathrm{i}\delta}
        -\frac{A_{n,m}^*(r,r^\prime)}{\omega+(\epsilon_m-\epsilon_n)-\mathrm{i}\delta
        }	\right),\hspace{6mm}\label{eq:P}
\end{eqnarray}
with
\begin{eqnarray}
        &&A_{n,m}(r,r^\prime) = \Psi_n^*(r)\Psi_n(r^\prime)\Psi_m(r)\Psi_m^*(r^\prime).\label{eq:P-cRPA}
\end{eqnarray}
Here, $\Psi_n$ is the $n$-th eigenstate of $H^0$ and $f$ is the 
Fermi function.  
In Fig.~\ref{fig:cRPA}, we show possible contributions to the sum in Eq.~(\ref{eq:P}).
Since the contribution to the polarization from transitions within the target subspace will be treated more accurately by solving the effective model,
the contribution denoted by (4) in Fig.~\ref{fig:cRPA} is excluded in the sum in Eq.~(\ref{eq:P}).
(The symbol $\sum^\prime$ means that these contributions are excluded.) 
Since $P$ is spin-diagonal and $v$ is spin-independent, Eq.~(\ref{eq:screening-rs}) reads
\begin{align}
  &w(\boldr\sigma,\boldr^\prime\sigma^\prime) \nonumber\\
  &= v(\boldr,\boldr^\prime)\!+\! \int \mathrm{d} \boldr_1 \mathrm{d} \boldr_2 \sum_{\sigma_1,\sigma_2}v(\boldr\sigma,\boldr_1\sigma_1) \delta_{\sigma_1\sigma_2}P(\boldr_1\sigma_1,\boldr_2\sigma_2) \nonumber \\
  & \quad \times w(\boldr_2\sigma_2,\boldr^\prime\sigma^\prime),\nonumber \\
  &= v(\boldr,\boldr^\prime)\!+\! 2\int \mathrm{d} \boldr_1 \mathrm{d} \boldr_2 v(\boldr,\boldr_1) P(\boldr_1,\boldr_2) w(\boldr_2,\boldr^\prime).\label{eq:screening-rs2}
\end{align}
The factor of 2 in the last line comes from the sum over two screening processes involving different spin sectors of $P$ , i.e., $P(\uparrow,\uparrow)$ and $P(\downarrow,\downarrow)$.

Once the screened two-body interaction has been computed, the screened interaction is projected onto the basis of the target manifold.
Replacing $v(\boldsymbol{r}-\boldsymbol{r}^\prime)$ with $w(\boldsymbol{r}-\boldsymbol{r}^\prime)$ in Eq.~(\ref{eq:Vijkl}),
the screened interaction is given by
\begin{eqnarray}
  \hat{W}(\omega) &=& \frac{1}{2}\sum_{ijkl} \sum_{\sigma_1\sigma_2\sigma_3\sigma_4} W_{ijkl}^{\sigma_1\sigma_2\sigma_3\sigma_4}(\omega) d^\dagger_{i\sigma_1} d^\dagger_{j\sigma_2} d_{k\sigma_3} d_{l\sigma_4}\label{eq:prj-W}.\hspace{2mm}
\end{eqnarray}
Here, $\hat{d}_{i\sigma}$ and $\hat{d}^\dagger_{i\sigma}$ are the annihilation and creation operators corresponding to Wannier orbitals of the target manifold, which will be constructed below.
The matrix $\boldsymbol{W}$ is given by
\begin{eqnarray}
  W_{ijkl}(\omega) &=& \int\diff r \diff r^\prime  \bar{\phi}^*_{i\sigma_1}(r) \bar{\phi}^*_{j\sigma_2}(r^\prime) w(\omega,\boldsymbol{r}-\boldsymbol{r}^\prime)\nonumber\\
  && \times \bar{\phi}_{k\sigma_3}(r^\prime) \bar{\phi}_{l\sigma_4}(r).\label{eq:Wijkl}
\end{eqnarray}
The $\omega$ dependence of $W$ can be 
accounted for in the solution of the effective model. 

A convenient way to solve the cRPA equation is to introduce the so-called product basis.~\cite{Aryasetiawan:1994hn}
Assuming that the orthonormal localized basis $\phi_i$ is real, that is, $\phi_i^*(\boldsymbol{r}) = \phi_i(\boldsymbol{r})$,
the product basis is defined by 
\begin{eqnarray}
  \{B_{ij}(\boldsymbol{r})\} = \{\phi_i(\boldsymbol{r})\phi_j(\boldsymbol{r})\}.\label{eq:product-basis-lattice}
\end{eqnarray}
Note that the product basis is not orthonormal.
In the following, we use $I$ and $I^\prime$ to refer to the index of the product basis, that is, $I\equiv (ij)$.
We expand $P(r,r^\prime)$ in terms of the product basis as
\begin{eqnarray}
  P(r,r^\prime) &=& \delta_{\sigma\sigma^\prime} P(\boldsymbol{r},\boldsymbol{r}^\prime)\nonumber \\
  &=& \delta_{\sigma\sigma^\prime} \sum_{I,I^\prime} P_{II^\prime}B_I(\boldsymbol{r}) B_{I^\prime}(\boldsymbol{r}^\prime).~\label{eq:Pmat}
\end{eqnarray}
Then, Eq.~(\ref{eq:screening-rs2}) and (\ref{eq:Wijkl}) lead to
\begin{align}
  & W_{ijkl} = W_{II^\prime} \nonumber\\
  &= \int\diff \boldsymbol{r} \diff \boldsymbol{r}^\prime B_I(\boldsymbol{r}) W(\boldsymbol{r}-\boldsymbol{r}^\prime) B_{I^\prime}(\boldsymbol{r}^\prime)\nonumber\\
  &= \int\diff \boldsymbol{r} \diff \boldsymbol{r}^\prime B_I(\boldsymbol{r}) V(\boldsymbol{r}-\boldsymbol{r}^\prime) B_{I^\prime}(\boldsymbol{r}^\prime) \nonumber \\
  & + 2P_{I_1I_2} \int \diff \boldr \diff \boldr_1 B_I(\boldr)V(\boldr-\boldr_1) B_{I_1} (\boldr_1)\nonumber \\
  & \quad\times \int \diff \boldr_2 \diff \boldr^\prime B_{I_2}(\boldr_2) W(\boldr_2-\boldr^\prime) B_{I^\prime}(\boldr^\prime)\nonumber \\
  &= V_{II^\prime} + 2\sum_{I_1I_2} V_{II_1} P_{I_1 I_2} W_{I_2I^\prime},
\end{align}
where we take $I=(ik)$ and $I^\prime=(jl)$.
This equation can be written in matrix representation as
\begin{align}
  W &= V + 2V P W  
   = (I - 2VP)^{-1} V.\label{eq:RPA-matrix}
\end{align}
We do not need to consider the spin degrees of freedom in Eq.~(\ref{eq:RPA-matrix}) since Eq.~(\ref{eq:screening-rs2}) is already spin-independent.

For our model, the product basis is given by 
\begin{eqnarray}
  \{B_I\}&=&\{\phi_{ia}^2\},
\end{eqnarray}
since we include only density-density interactions.
In other words, the integral in Eq.~(\ref{eq:Vijkl}) vanishes whenever terms like $\phi_{ia}(\boldsymbol{r})\phi_{jb}(\boldsymbol{r})$ ($i\neq j$ or $a\neq b$) appear.
The index $a$ denotes orbital and does not include spin.

We define the Fourier transformations of the bare Coulomb interaction $V$ and the polarization $P$ as
\begin{eqnarray}
  V_{ab}(\boldsymbol{q}) &=& \frac{1}{N_k} \sum_i U_{ij}^{ab} e^{-\mathrm{i}\boldsymbol{q}\cdot (R_i-R_j)} = \frac{1}{N_k} U^{ab}, \\
  P_{ab}(\boldsymbol{q})&=& \frac{1}{N_k} \sum_i P_{ij}^{ab} e^{-\mathrm{i}\boldsymbol{q} \cdot (R_i-R_j)}. 
\end{eqnarray}
Diagonalizing the Fourier transformation of the one-body Hamiltonian, one obtains Bloch wavefunctions:
\begin{eqnarray}
  \Psi_{ka}(r) &=& \frac{1}{\sqrt{N_k}} \sum_{i} c_{kn,a} \phi_{ia}(r) e^{\mathrm{i}k R_i},\nonumber\\
  &=& \frac{1}{\sqrt{N_k}} \sum_{i} 
  \left(
  \begin{array}{c}
    c_{kn,1}\\
    c_{kn,2}\\
    \vdots\\
    c_{kn,N}\\
  \end{array}
  \right)
  e^{\mathrm{i}k R_i},
\end{eqnarray}
where $n$ is the band index.
Substituting this equation into Eq.~(\ref{eq:P-cRPA}), and using
\begin{eqnarray}
	&&P(r,r^\prime;\omega)= \sum_{n,m}^\prime f(\epsilon_n)(1-f(\epsilon_m))\times\nonumber \\
        &&\left(
   	\frac{A(r,r^\prime)}{\omega-(\epsilon_m-\epsilon_n)+\mathrm{i}\delta}
	-\frac{A^*(r,r^\prime)}{\omega+(\epsilon_m-\epsilon_n)-\mathrm{i}\delta
        }	\right),~\label{eq:P-2nd}
\end{eqnarray}
we obtain 
\begin{eqnarray}
  P_{a b}(\boldsymbol{q};\omega) &=& 
    \frac{1}{N_k}
    \sum_{k n n^\prime}^\prime f_{k,n} (1-f_{k+q,n^\prime})\nonumber\\
    && \times \Bigg[ \frac{c^*_{kn}(a) c_{k+q n^\prime}(a) c_{kn}(b) c^*_{k+q n^\prime}(b)}{\omega-(\epsilon_{k+q,n^\prime}-\epsilon_{k,n})+\mathrm{i}\delta}\nonumber \\
    && -\frac{c_{kn}(a) c^*_{k+q n^\prime}(a) c^*_{kn}(b) c_{k+q n^\prime}(b)}{\omega+(\epsilon_{k+q,n^\prime}-\epsilon_{k,n})-\mathrm{i}\delta} \Bigg],\hspace{2mm}
\end{eqnarray}
where $\epsilon_{k,n}$ is the $n$-th eigenvalue at wavevector $k$.

In reciprocal space, the cRPA equation reads
\begin{eqnarray}
  W_{ab}(\boldsymbol{q}) &=& V(\boldsymbol{q})_{ab}+2 \sum_{cd} V_{a c}(\boldsymbol{q}) P_{cd}(\boldsymbol{q}) W_{db}(\boldsymbol{q}),
\end{eqnarray}
where $a,b,c,d$ are orbital indices.
This can be rewritten in matrix form as
\begin{eqnarray}
  \boldsymbol{W}(\boldsymbol{q}) &=& [\boldsymbol{I} - 2\boldsymbol{V}(\boldsymbol{q})\boldsymbol{P}(\boldsymbol{q})]^{-1} \boldsymbol{V(\boldsymbol{q})}.\label{eq:cRPA-q}
\end{eqnarray}

Next, $\boldsymbol{W}(\boldsymbol{q})$ is projected onto a localized basis for the target band(s).
A set of Wannier functions localized in the unit cell $\boldsymbol{R}_i$ is defined by
\begin{eqnarray}
  |\boldsymbol{R}_i n\rangle &=& \frac{1}{N_k} \sum_{\boldsymbol{k}} \left(\sum_m {\mathcal U}_{mn}^{\boldsymbol{k}} e^{-\mathi \boldsymbol{k}\cdot \boldsymbol{R}} |\Psi_{m\boldsymbol{k}}\rangle\right)\nonumber\\
  &=& \sum_j \alpha_{ija}^n \phi_{ja}(\boldr),\label{eq:wannier}
\end{eqnarray}
where $n=1,\ldots, N_\text{band}$ is the index of the Wannier function in a unit cell ($N_\text{band}$ is the number of target bands).
The symbol $a$ denotes the orbital index in the unit cell $j$.
When we construct maximally localized Wannier functions~\cite{Souza:2001ba,Marzari:329970} from Bloch wavefunctions obtained by first-principles calculations,
the gauge matrix $\mathcal{U}_{mn}^{\boldsymbol{k}}$ is chosen so that the Wannier functions are localized in real space.

To obtain $\mathcal{U}_{mn}^{\boldsymbol{k}}$ we write the non-interacting part of the Hamiltonian of our model as
\begin{align}
  \boldsymbol{H}(\boldsymbol{k}) &=  -2t\left(\cos(k_x)+\cos(k_y)+\cos(k_z)\right) \boldsymbol{I} + \boldsymbol{H}_0,
\end{align}
where 
\begin{eqnarray}
  \boldsymbol{H}_0 &=& \begin{pmatrix}
      -\Delta & -t^\prime & 0 \\
      -t^\prime & 0 & -t^\prime \\
      0  & -t^\prime & \Delta 
    \end{pmatrix}.
\end{eqnarray}
Since the Bloch wavefunction is independent of wavevector,
we can take a unitary matrix $\boldsymbol{\mathcal U}=\left\{ \boldsymbol{u}_1,\boldsymbol{u}_2, \boldsymbol{u}_3\right\}$
that diagonalizes $\boldsymbol{H}_0$,
and denote the eigenvalues by $\epsilon_1,\epsilon_2,\epsilon_3$.
This also diagonalizes the full non-interacting Hamiltonian at the same time,
and the eigenvalues are $-2t\cos(\boldsymbol{k})+ \epsilon_1,-2t\cos(\boldsymbol{k})+ \epsilon_2 , -2t\cos(\boldsymbol{k})+ \epsilon_3$.
Taking the gauge matrix $\boldsymbol{\mathcal U}(\boldsymbol{k})=\boldsymbol{I}$,
the Wannier function for the target band localized at site $i_0$ becomes 
\begin{eqnarray}
  |i;i_0\rangle &=& \delta_{i i_0} \boldsymbol{u}_2,
\end{eqnarray}
where $i$ is the site index.

Next we project the screened interactions onto the Wannier basis of the target band.
The matrix elements in Eq.~(\ref{eq:prj-W}) have non-zero values only when ($i=l$ and $j=k$) and ($\sigma_2=\sigma_3$ and $\sigma_1=\sigma_4$).
From Eq.~(\ref{eq:prj-W}), we obtain
\begin{eqnarray}
  \hat{W} &=& U \sum_i \hat{n}_{i\uparrow} \hat{n}_{i\downarrow} + \frac{1}{2} \sum_{i\neq j} V_{ij}\hat{n}_i\hat{n}_j,\hspace{5mm}\label{eq:eff-int-gen}
\end{eqnarray}
where 
\begin{eqnarray}
  U &=& \sum_{ab} W(\boldsymbol{R}=0)_{ab} |u_2(a)|^2|u_2(b)|^2,\\
  V_{ij} &=& \sum_{ab} W(\boldsymbol{R}_i -\boldsymbol{R}_j)_{ab} |u_2(a)|^2|u_2(b)|^2.\hspace{3mm}
\end{eqnarray}
Exchange integrals vanish since the orbitals $\phi_{ia}$ are taken to be delta functions in our model.

\subsection{Pauli principle and spin-dependent formalism}\label{sec:spin-dep-cRPA}
The RPA method violates the Pauli principle for a Hubbard-like model
because its diagrammatic expansion contains self interactions between same-spin electrons. 
To remove diagrams violating the Pauli principle, we introduce a spin-dependent formalism.
This idea is similar to the self-interaction correction for the GW method.~\cite{Aryasetiawan:2012dta}
Restricting ourselves to density-density interactions, we consider the product basis 
\begin{eqnarray}
  \{B_{I}\}&=&\{\phi_{ia\sigma}^2(r)\},
\end{eqnarray}
similarly to Eq.~(\ref{eq:product-basis-lattice}).
Here $\sigma$ is the spin quantum number and $r$ is the composite index of spin and position.
Note that $B_{I\sigma}(r)$ is nonzero only for the spin sector $\sigma$.
The bare Coulomb matrix $\boldsymbol{V}$ is given by
\begin{eqnarray}
  V_{I\sigma,I^\prime\sigma^\prime} &=& \int dr dr^\prime B_{I\sigma}(r) V(r-r^\prime) B_{I^\prime\sigma^\prime}(r^\prime).
\end{eqnarray}
Following the Pauli principle, the Coulomb matrix is now taken to be spin dependent.
In other words, $V_{ I\sigma,I^\prime\sigma^\prime}=0$ for $I=I^\prime$ and $\sigma=\sigma^\prime$.

After Fourier transformation, the cRPA equation reads
\begin{eqnarray}
  \boldsymbol{W}(\boldsymbol{q}) &=& [\boldsymbol{I} - \boldsymbol{V}(\boldsymbol{q})\boldsymbol{P}(\boldsymbol{q})]^{-1} \boldsymbol{V(\boldsymbol{q})}.\label{eq:spin-cRPA-q}
\end{eqnarray}
Note that the factor of 2 in front of the $\boldsymbol{V}$ in Eq.~(\ref{eq:RPA-matrix}) is no more needed for the spin-dependent formalism because the summation over spin is taken into account by the matrix formalism.
The polarization function $\boldsymbol{P}$ is spin-diagonal and spin-independent, and its matrix elements are given by Eq.~(\ref{eq:Pmat}).

To see how the two formalisms give different results for on-site repulsions,
let us consider a simplified version of the three-orbital model introduced in Sec.~\ref{sec:model}.
We take $U_1=U_3=0$.
For the spin-independent formalism, the bare Coulomb matrix is a 3$\times$3 matrix defined as
\begin{eqnarray}
  \boldsymbol{V}(\boldsymbol{q}) &=&
        \begin{pmatrix}
         0        & 0         & 0  \\
         0 & U_d       & 0  \\
         0 & 0  & 0   \\
        \end{pmatrix}.\\
\end{eqnarray}
Expanding the cRPA equation with respect to $U_d$,
we obtain the screened interaction projected on the orbital 2 as
\begin{eqnarray}
    W_{22}(\boldsymbol{q}) &=& V_{22} + 2V_{22} P_{22}(\boldsymbol{q}) V_{22} +  O(P^2)\\
    &=& U_d + 2 U_d P_{22}(\boldsymbol{q}) U_d +  O(P^2).\label{eq:W22}
\end{eqnarray}
However, the second term in the last line should not exist due to the Pauli principle.
In other words, the expansion must start from the second-order term, i.e, $O(P^2)$ because
the on-site repulsion acts only between the up-spin and down-spin sectors.
This constraint is missing in the spin-independent cRPA procedure.

On the other hand,
in the spin-dependent formalism, the Coulomb matrix is a $6\times 6$ matrix of the form
\begin{eqnarray}
  \boldsymbol{V}(\boldsymbol{q}) &=&
    \begin{pmatrix}
      \boldsymbol{V}_{\uparrow\uparrow} & \boldsymbol{V}_{\uparrow\downarrow} \\
      \boldsymbol{V}_{\downarrow\uparrow} & \boldsymbol{V}_{\downarrow\downarrow}
    \end{pmatrix},
\end{eqnarray}
where
\begin{eqnarray}
  \boldsymbol{V}_{\uparrow\uparrow}(\boldsymbol{q}) &=& \boldsymbol{V_{\downarrow\downarrow}}(\boldsymbol{q})=
    \begin{pmatrix}
     0        & 0         & 0 \\
     0        & 0         & 0 \\
     0        & 0         & 0        \\
    \end{pmatrix},\\
    \boldsymbol{V}_{\uparrow\downarrow}(\boldsymbol{q}) &=& \boldsymbol{V}_{\downarrow\uparrow}(\boldsymbol{q})=
        \begin{pmatrix}
     0        & 0         & 0 \\
     0        & U_d         & 0 \\
     0        & 0         & 0        \\
      \end{pmatrix}.
\end{eqnarray}
The screened interaction projected on the orbital 2 is now 
\begin{eqnarray}
    W_{2\uparrow2\downarrow}(\boldsymbol{q}) &=& V_{2\uparrow2\downarrow} + V_{2\uparrow2\downarrow} P_{22}(\boldsymbol{q}) V_{2\downarrow 2\uparrow} P_{22}(\boldsymbol{q}) V_{2\uparrow 2\downarrow} \nonumber\\
    && + O(P^4)\\
   &=& U_d + U_d^3 P_{22}^2(\boldsymbol{q}) + O(P^4).\label{eq:W22-spin}   
\end{eqnarray}
Comparing Eqs.~(\ref{eq:W22}) and (\ref{eq:W22-spin}), we immediately see that the Pauli principle is restored in the spin-dependent cRPA formalism.
This difference can be substantial if the target manifold is strongly correlated.

A drawback is that this formalism breaks the $SU(2)$ symmetry of the full model because we ignore transverse spin susceptibilities.
Restoring this symmetry within the RPA formalism is nontrivial and we will not attempt this here.

\section{Dynamical mean-field theory}\label{sec:dmft}
We use variants of the dynamical mean-field theory to analyze the full multi-orbital models and the effective one-band models in a consistent way.
We use multi-orbital DMFT for the full models.
The effective models are solved using the extended DMFT framework which can treat off-site interactions within a single-impurity description.

\subsection{Multi-orbital DMFT}\label{sec:dmft}
In order to solve the full multi-orbital model, we use DMFT.
Its self-consistency loop is given by
\begin{eqnarray}
  \boldsymbol{\Sigma}(\mathrm{i}\omega_n) &=& \boldsymbol{\mathcal{G}}(\mathrm{i}\omega_n)^{-1}-\boldsymbol{G}_\mathrm{imp}^{-1}(\mathrm{i}\omega_n), \label{eq:self-energy}\\
  \boldsymbol{G}_\mathrm{loc}(\mathrm{i}\omega_n) &=& \frac{1}{N_k}\sum_k \frac{1}{\mathi\omega_n + \mu - \mathcal{H}_0(k)  - \boldsymbol{\Sigma}(\mathrm{i}\omega_n)},\hspace{1mm}\label{eq:Gloc}\\
  \boldsymbol{\mathcal{G}}(\mathrm{i}\omega_n)^{-1} &=& \boldsymbol{G}_\mathrm{loc}^{-1}(\mathrm{i}\omega_n) + \boldsymbol{\Sigma}(\mathrm{i}\omega_n),\label{eq:cavity}
\end{eqnarray}
where $\mathcal{H}_0(\boldsymbol{k})$ is the Fourier transform of the one-body part of the Hamiltonian. 
$\boldsymbol{\Sigma}$, $\boldsymbol{G}_\mathrm{loc}$, and $\boldsymbol{\mathcal{G}}$ 
are the self-energy, local Green's function, and the Weiss function, 
respectively. 
Since we consider the paramagnetic case, they are $N_\mathrm{orb} \times N_\mathrm{orb}$ matrices.

After obtaining $\boldsymbol{\mathcal{G}}$ in Eq.~(\ref{eq:cavity}), we solve the multi-orbital quantum impurity problem given by the action
\begin{eqnarray}
  S &=& - \sum_{ab\sigma} \int_0^\beta \diff \tau \diff \tau^\prime  c^\dagger_{a\sigma}(\tau) \mathcal{G}_{ab}^{-1}(\tau-\tau^\prime) c_{b\sigma}(\tau^\prime) \nonumber\\
  && +\frac{1}{2} \sum_{ab}\int_0^\beta \diff \tau U_{ab} n_a(\tau) n_b(\tau),
\end{eqnarray}
where $U_{ab}$ is the on-site part of the density-density interaction.
We employ a continuous-time quantum Monte Carlo impurity solver based on the hybridization expansion and the matrix formalism.~\cite{Werner:2006ko,Werner:2006iz}
The sign problem is reduced by rotating the basis of the hybridization function.
We refer to Appendices~\ref{appendix:basis-rot} for details.

After computing $\boldsymbol{G}$, $\boldsymbol{\Sigma}$ is updated using Eq.~(\ref{eq:self-energy}) and the self-consistency loop is repeated until a converged solution is obtained.

In our analyses, we project the Green's function $\boldsymbol{G}(\mathi\omega_n)$ onto the basis that diagonalizes $\langle H(k)\rangle_k$ as
\begin{eqnarray}
  G_m(\mathi\omega_n) &=& \boldsymbol{u}_m^\dagger \boldsymbol{G}(\mathi\omega_n) \boldsymbol{u}_m,\label{eq:prj-Gomegan}
\end{eqnarray}
where $\boldsymbol{u}_m$ is the $m$-th eigenvector of $\langle H(k)\rangle_k$.
The Green's function is defined by
\begin{eqnarray}
  \boldsymbol{G}_{ab}(\tau) &=& -\langle T_\tau c_a(\tau) c^\dagger_b(0) \rangle,\\
  \boldsymbol{G}_{ab} (\mathi\omega_n) &=& \int_0^\beta \mathrm{d}\tau e^{\mathi\omega_n \tau} \boldsymbol{G}(\tau),
\end{eqnarray}
where $T_\tau$ denotes imaginary-time ordering and $\omega_n = (2n+1) \pi/\beta$.
We call this basis the ``band basis''.
The renormalization factor $Z$ is computed by using the approximation
\begin{eqnarray}
  Z&=&\frac{1}{1-\frac{\partial \Sigma_m(\omega)}{\partial \omega}} \approx \frac{1}{1-\frac{\text{Im}\Sigma_m(\mathi\omega_0)}{\pi/\beta}},~\label{eq:Z}
\end{eqnarray}
where $m$ is the index of the target band.

\subsection{EDMFT}\label{sec:edmft}
To solve the effective one-band model with dynamical on-site and off-site interactions,
we use extended DMFT (EDMFT).~\cite{Ayral:2013ca,Si:1996wh,Sengupta:1995wy,Sun:2002jb,Sun:2004dm}
This formalism can treat off-site interactions, even though it is based on a single-site impurity construction.
In the present study, we consider only dynamical nearest-neighbor interactions.
In the EDMFT calculation, we have to solve the impurity problem  
\begin{eqnarray}
   S &=& -\sum_{ab\sigma}\int_0^\beta \diff \tau \diff \tau^\prime c^\dagger_{a\sigma}(\tau) \mathcal{G}^{-1}_{ab}(\tau-\tau^\prime) c_{b\sigma}(\tau^\prime) \nonumber\\
  && + \int_0^\beta \diff \tau \diff \tau^\prime  n^\dagger(\tau) U(\tau-\tau^\prime) n(\tau^\prime).
\end{eqnarray}
The retarded interaction $U(\tau)$ is determined by the following self-consistency equations, which are similar to Eqs.~(\ref{eq:Gloc})--(\ref{eq:self-energy}): 
\begin{eqnarray}
  W_\text{imp}(\mathrm{i}\nu_n) &=& {\mathcal{U}}(\mathrm{i}\nu_n) - {\mathcal{U}}(\mathrm{i}\nu_n)  \chi_\text{imp}(\mathrm{i}\nu_n)  {\mathcal{U}}(\mathrm{i}\nu_n),\\
  \Pi(\mathrm{i}\nu_n) &=& U(\mathrm{i}\nu_n)^{-1}-W^{-1}_\text{imp}(\mathrm{i}\nu_n),\\
  {W}_\mathrm{loc}(\mathrm{i}\nu_n) &=& \frac{1}{N_k}\sum_k \frac{1}{{v}_k^{-1}(\mathi \nu_n) - {\Pi}(\mathrm{i}\nu_n)},\\
  U(\mathrm{i}\nu_n)^{-1} &=& W^{-1}_\text{loc}(\mathrm{i}\nu_n) + \Pi(\mathrm{i}\nu_n),
  \end{eqnarray}
where 
\begin{eqnarray}
  v_k(\mathi \nu_n) &=&  \sum_i v_i(\mathi \nu) e^{\mathi r_i k},\\
  \chi_\mathrm{imp}(\tau) &=& \langle n(\tau)n(0) \rangle - \langle n \rangle^2.
\end{eqnarray}
This impurity problem with retarded density-density interaction is also solved by the hybridization expansion method.~\cite{Werner:2006iz,Werner:2006ko,Werner:2007ix,Werner:2010cd}
Note that the dynamical nature of the screened interactions causes a renormalization of the kinetic energy and band width.~\cite{Casula:2012ba}
This effect will be taken into account in solving our low-energy models by treating the frequency dependence of the screened interactions explicitly within the EDMFT framework.
For the effective model obtained by the spin-cRPA method, we take into account only retarded on-site interaction and ignore longer ranged interactions.
This point will be discussed again in Sec.~\ref{sec:results-cRPA}.
  
\section{Results}
\subsection{Downfolded models}\label{sec:results-cRPA}
\begin{figure*}
 \begin{tabular}{c}
 \begin{minipage}{0.5\hsize}
  \centering
  \includegraphics[width=0.95\textwidth,clip,type=pdf,ext=.pdf,read=.pdf]{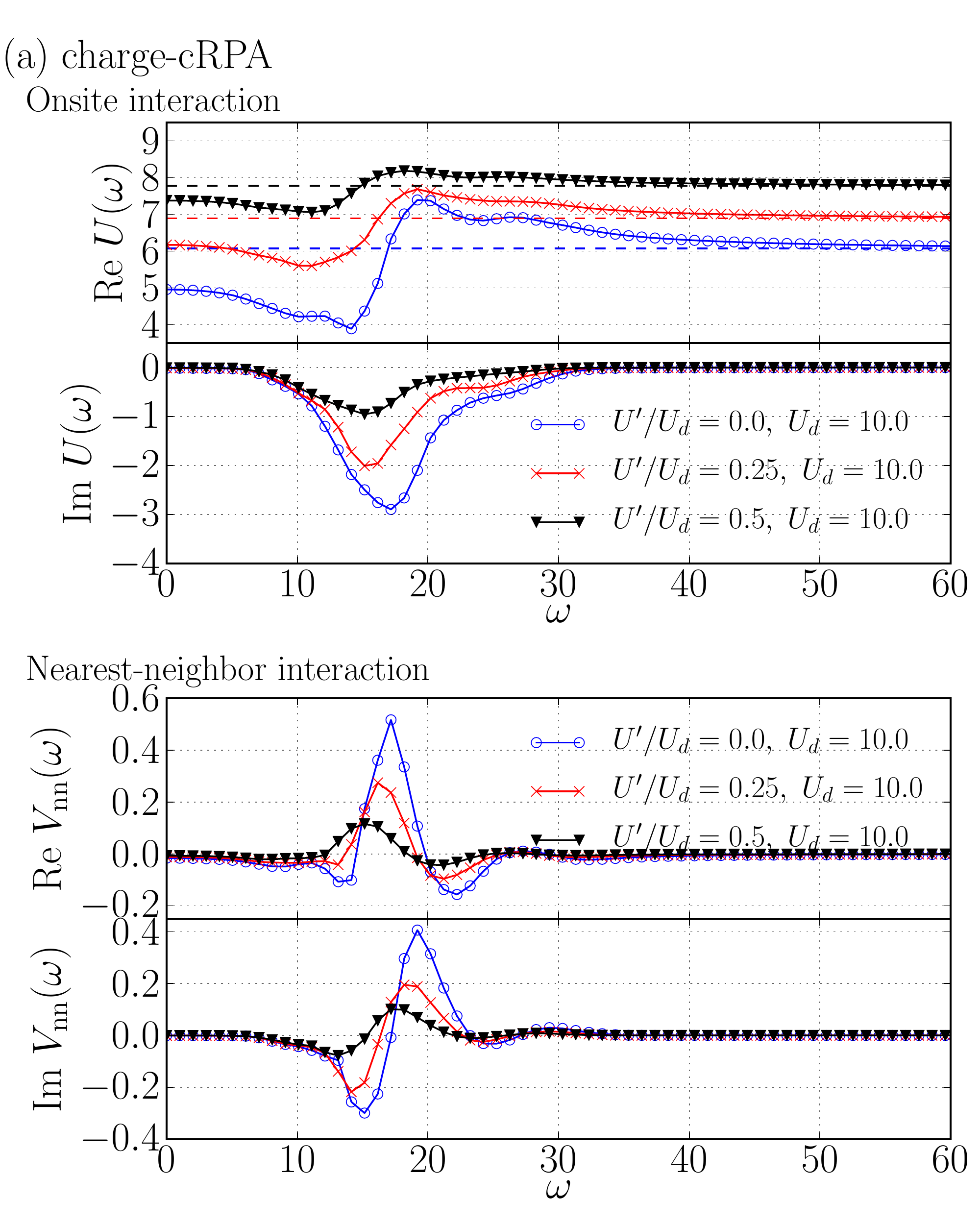}
 \end{minipage}
 \begin{minipage}{0.5\hsize}
  \centering
  \includegraphics[width=0.95\textwidth,clip,type=pdf,ext=.pdf,read=.pdf]{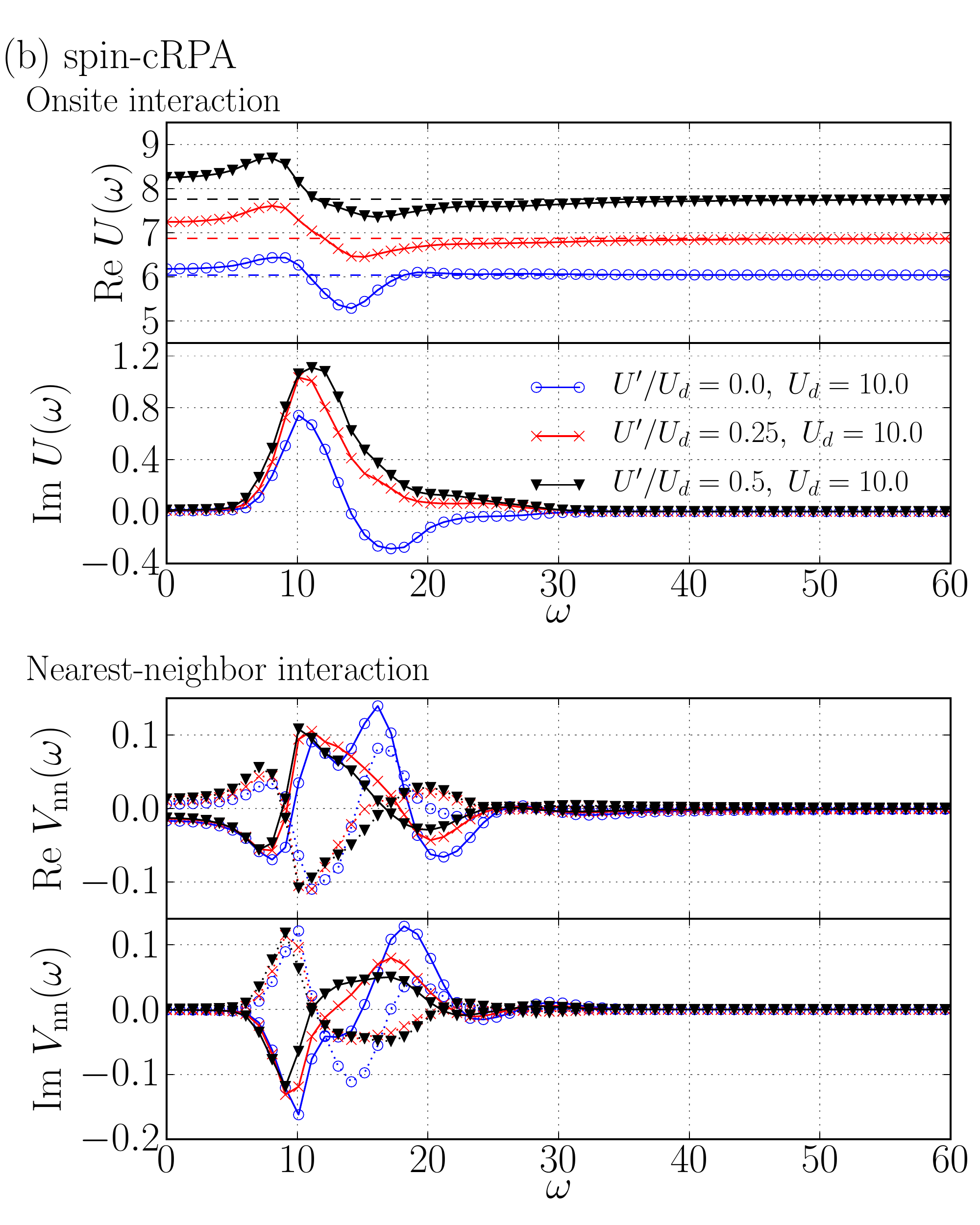}
 \end{minipage}
 \end{tabular}
 \caption{
 (Color online)
 Screened interactions obtained by the charge-cRPA [(a)] and the spin-cRPA [(b)] for the three-orbital model ($\Delta=10$ and $t^\prime=4$).
 We show the on-site interaction $U(\omega)$ and nearest neighbor interaction $V_\mathrm{nn} (\omega)$.
 The bare on-site interaction $U(\omega=\infty)$ is represented by a horizontal solid line.
 For the nearest neighbor interactions computed by the spin-cRPA [(b)], the solid and broken lines represent the spin-diagonal element of the interaction [$V_\mathrm{nn}^{\uparrow\uparrow}$~( $=V_\mathrm{nn}^{\downarrow\downarrow}$)] and the spin off-diagonal one ($V_\mathrm{nn}^{\uparrow\downarrow}$), respectively. 
 }
 \label{fig:3orb-scr-int-omega-dep}
\end{figure*}
\begin{figure}
 \centering\includegraphics[width=.5\textwidth,clip,type=pdf,ext=.pdf,read=.pdf]{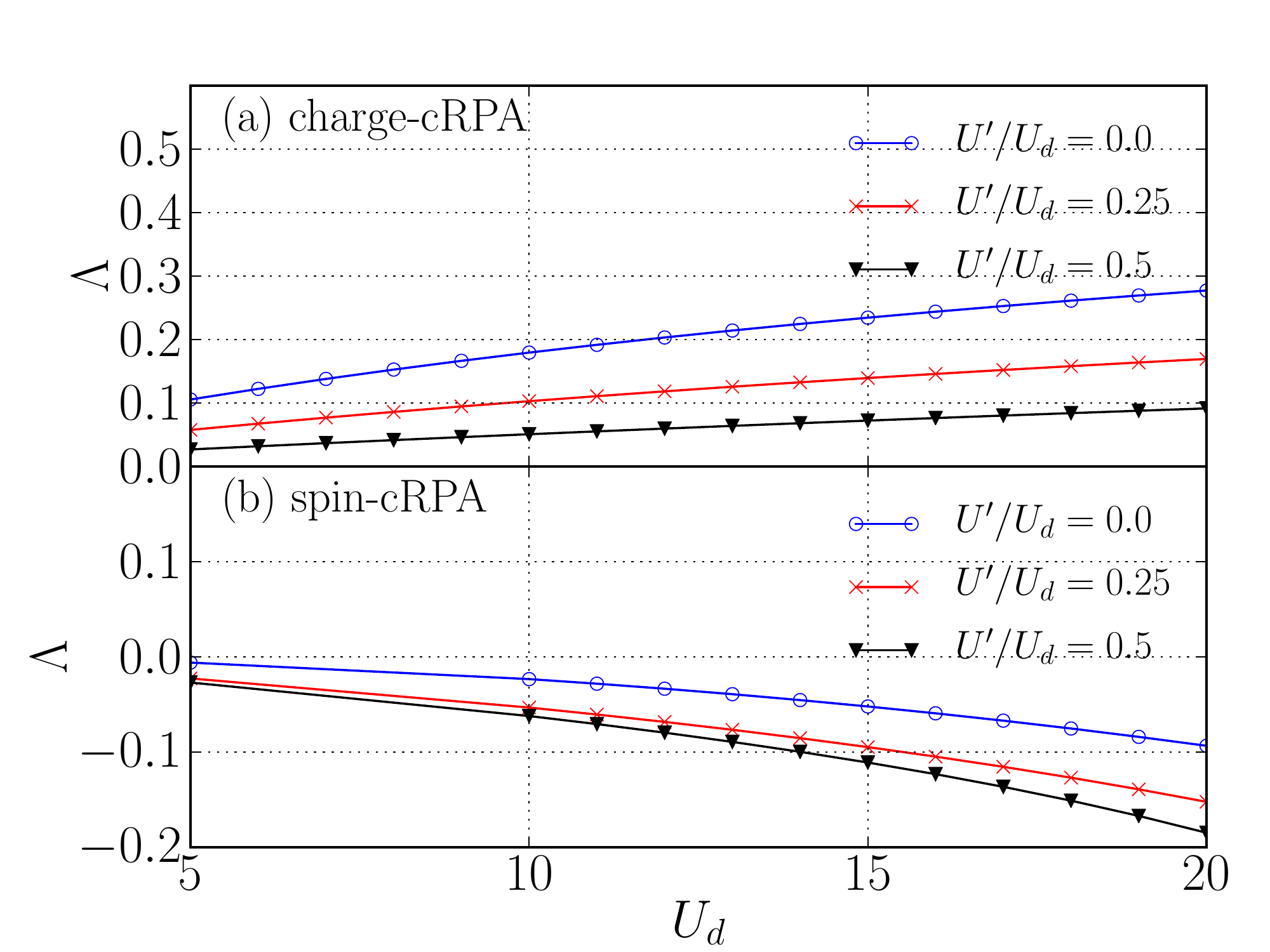}
 \caption{
 (Color online)
 Strength of the screening effect for the on-site interaction $\Lambda$ computed by the charge-cRPA [(a)] and spin-cRPA [(b)] methods for the three-orbital model.
 The definition of $\Lambda$ is given in Eq.~(\ref{eq:strength-screening}).
 }
 \label{fig:screening}
\end{figure}

We now derive low-energy effective models for the three-orbital model using the cRPA method.
Figures~\ref{fig:3orb-scr-int-omega-dep}(a) and (b) compare the screened interactions computed by the charge-cRPA and spin-cRPA methods.
We show the on-site interaction $U(\omega)$ and nearest-neighbor interaction $V_\mathrm{nn}(\omega)$ for $U_d$=10 and typical values of $U^\prime$.
These parameter sets correspond to the correlated metal phase (see the phase diagram in Fig.~\ref{fig:3orb-gap10-pd}).
Let us first look at the results by the charge-cRPA method.
For all the values of $U^\prime/U_d$ considered here,
Im$U(\omega)$ exhibits two negative peaks located around $\omega$=15 and $\omega$=25.
Below these energy scales, the on-site interaction is reduced from the instantaneous value $U(\omega=\infty)$.
Those two energy scales correspond to transitions between the target band and the screening bands, and those between the lower and upper target bands, respectively.
The peak at the smaller $\omega$ is higher than the other one,
indicating that the former contribution dominates in the screening effects. 
We also note that $U(\omega=\infty)$ is smaller than $\Ud$ because the Wannier function extends to the less correlated screening orbitals.
Although the full model has only on-site interactions,
a dynamic nearest-neighbor interaction $V_\mathrm{nn}(\omega)$ is generated.
This interaction is substantially smaller than the on-site interaction, and almost vanishes at low frequencies.
The full $\omega$ dependence of the nearest-neighbor interaction is taken into account in the following EDMFT calculations.

We now move to the results obtained by the spin-cRPA method, which are displayed in Fig.~\ref{fig:3orb-scr-int-omega-dep} (b).
One immediately sees that the $\omega$ dependence of $U(\omega)$ is qualitatively different from the result obtained by the charge-cRPA method.
The two-peak structure in Im$U(\omega)$ is more apparent, and the screening frequencies are lower. 
A more substantial difference is that the first peak, which is associated with transitions between the target and screening bands, is positive in the case of the spin-cRPA method.
This contribution dominates over the other one, producing an anti-screening effect. 
The nearest-neighbor interaction is now spin dependent and has spin-diagonal and spin off-diagonal elements.
Although they have different $\omega$ dependences, the peaks in Im$V_\mathrm{nn}$ are substantially smaller than those in Im$U(\omega)$.

To quantify the strength of the screening of the on-site interaction, we evaluate 
\begin{equation}
  \Lambda \equiv 1-\frac{\mathrm{Re}U(\omega=0)}{\mathrm{Re}U(\omega=\infty)}\label{eq:strength-screening}
\end{equation}
for different $U^\prime$ and $U_d$.
$\Lambda>0$ correspond to a situation where the static interaction is screened. 
The result obtained by the charge-cRPA method is shown in Fig.~\ref{fig:screening}(a) as a function of $\Ud$.
Two notable trends are discernible.
First, $\Lambda$ becomes larger in the strongly correlated regime, that is, as $\Ud$ increases.
Second, $\Lambda$ increases if $U^\prime/U_d$ decreases.
The results obtained by the spin-cRPA are presented in Fig.~\ref{fig:screening}(b).
In this case, one always finds an anti-screening effect in the parameter regime considered.
The anti-screening effects become enhanced as $U^\prime$ is increased. 
Another notable point is that the nearest-neighbor interaction becomes spin-dependent in the spin-cRPA method.
This is because the spin-cRPA method breaks the $SU(2)$ symmetry.
To avoid this problem, we take into account only the on-site interaction in the DMFT calculations for the effective model obtained by the spin-cRPA method.

\subsection{DMFT results}
We analyze the three-orbital full model and the downfolded models within the DMFT or EDMFT framework.
In particular, we compare DMFT solutions of the following models:
\begin{itemize}
 \item  Single-band model (charge-cRPA),
 \item  Single-band model (spin-cRPA),
 \item  Single-band model with bare interactions $U(\omega)=U(\omega=\infty)$, $V_\mathrm{nn}(\omega)=V_\mathrm{nn}(\omega=\infty)$,
 \item  Single-band model (spin-cRPA) with renormalized band width (following Ref.~\onlinecite{Casula:2012ba}) and static interactions $U(\omega)=U(\omega=0)$, $V_\mathrm{nn}(\omega)=V_\mathrm{nn}(\omega=0)$,
 \item  Full three-orbital model.
\end{itemize}
In the following, we will refer to these models as the {\it charge-cRPA/spin-cRPA model}, the {\it unscreened model}, the {\it full model}, {\it spin-cRPA static model}, respectively.
The simulations are carried out at $\beta=15$ unless otherwise stated. 
We confirmed that this temperature is low enough to see ground-state behavior, i.e., the quasi-particle weights are essentially converged to the ground-state values.

\begin{figure}
 \centering\includegraphics[width=.4\textwidth,clip]{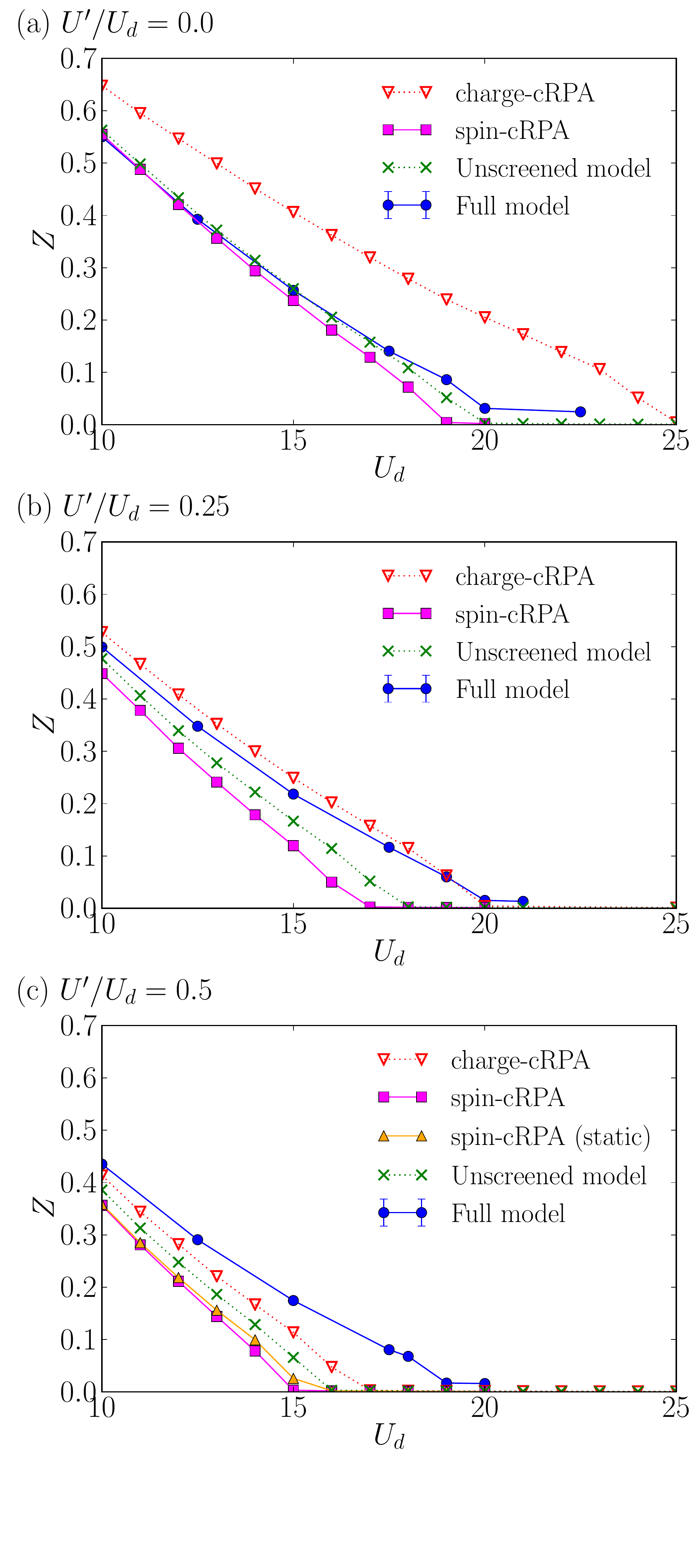}
 \caption{
 (Color online)
 Quasi-particle weights computed for the three-orbital model with $\Delta=10$ and $t^\prime=4$.
 We compare the results of the effective models downfolded by the charge-cRPA and spin-cRPA methods, the unscreened model, and the full model.
 }
 \label{fig:3orb-gap10-z}
\end{figure}

\begin{figure}
   \centering
   \includegraphics[width=0.35\textwidth,clip,type=pdf,ext=.pdf,read=.pdf]{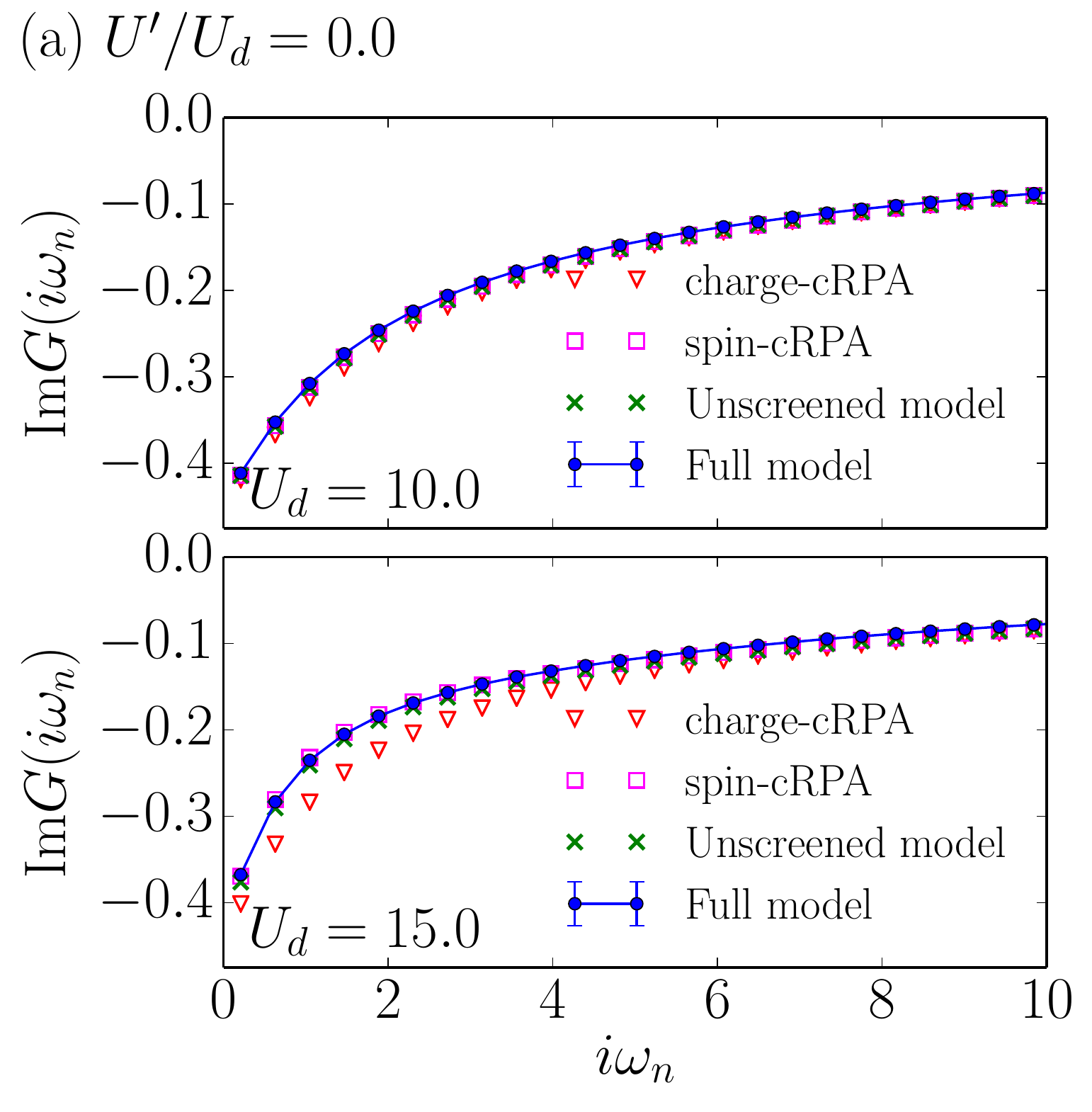}
   \includegraphics[width=0.35\textwidth,clip,type=pdf,ext=.pdf,read=.pdf]{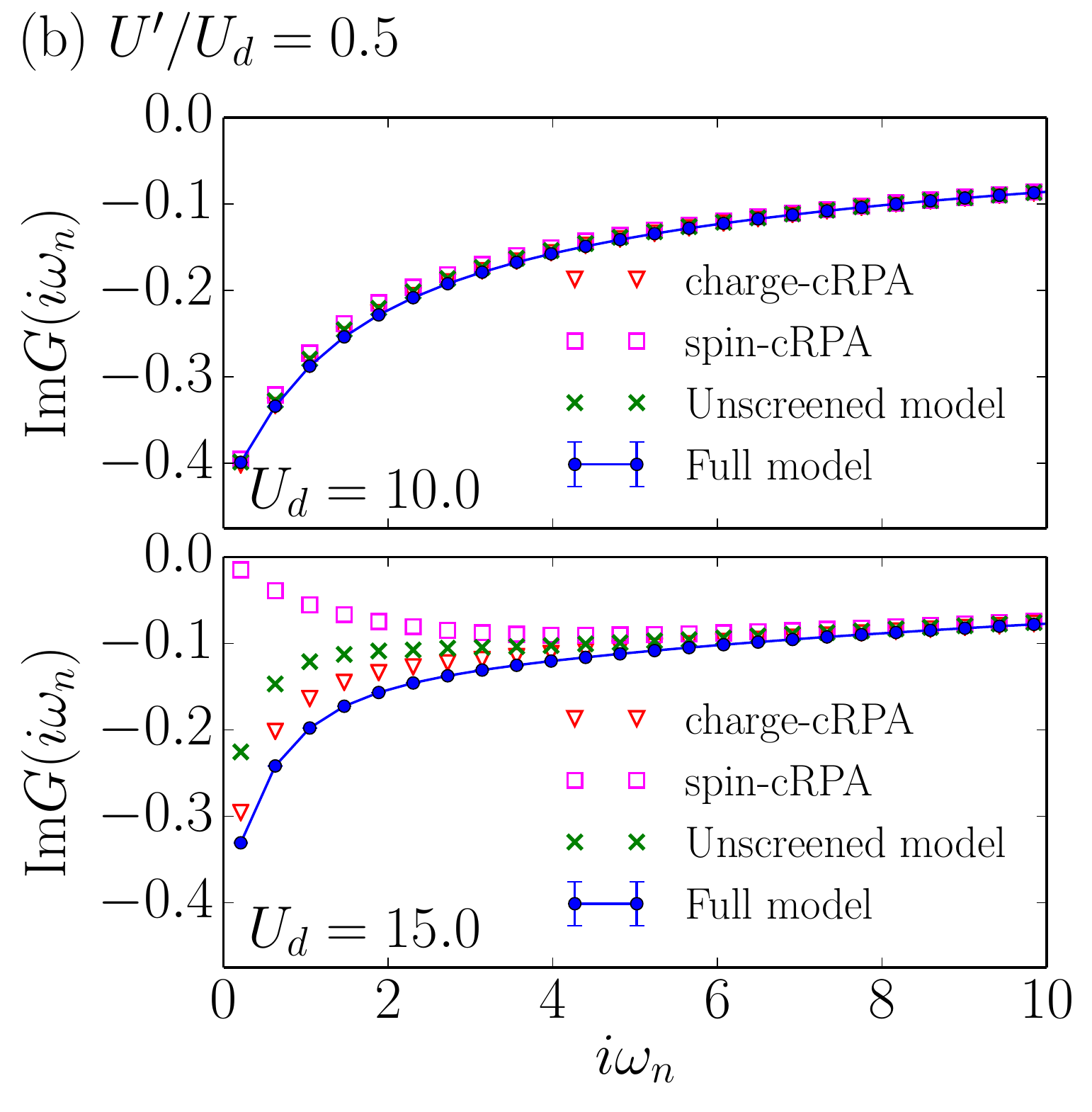}
 \caption{
 (Color online)
 Comparison of the local Green's function for the three-orbital model with $\Delta=10$ and $t^\prime=4$ ($\beta=15$).
 The triangles, squares, crosses filled circles denote the data obtained by solving the charge-cRPA model, the spin-cRPA model, and the unscreened model, and the full model, respectively.
 }
 \label{fig:3orb-gap10-g-comp}
\end{figure}

\begin{figure}
 \centering\includegraphics[width=.5\textwidth,clip,type=pdf,ext=.pdf,read=.pdf]{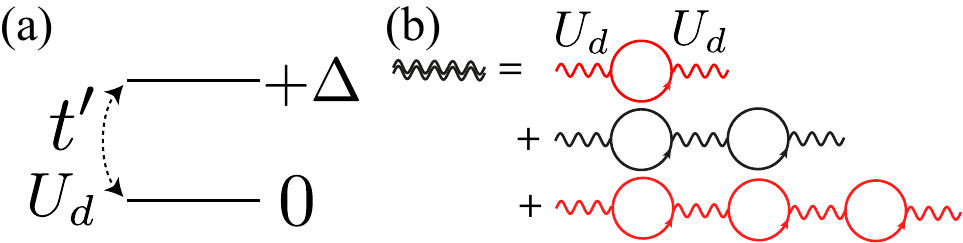}
 \caption{
 (Color online)
 (a) Two-orbital model with a lower target band and an upper screening band.
 (b) First three diagrams in the charge-cRPA series for the two-orbital model.
 The diagrams with an odd number of bubbles violate the Pauli principle. 
 }
 \label{fig:two-orbital-model}
\end{figure}

\begin{figure}
 \centering\includegraphics[width=.5\textwidth,clip,type=pdf,ext=.pdf,read=.pdf]{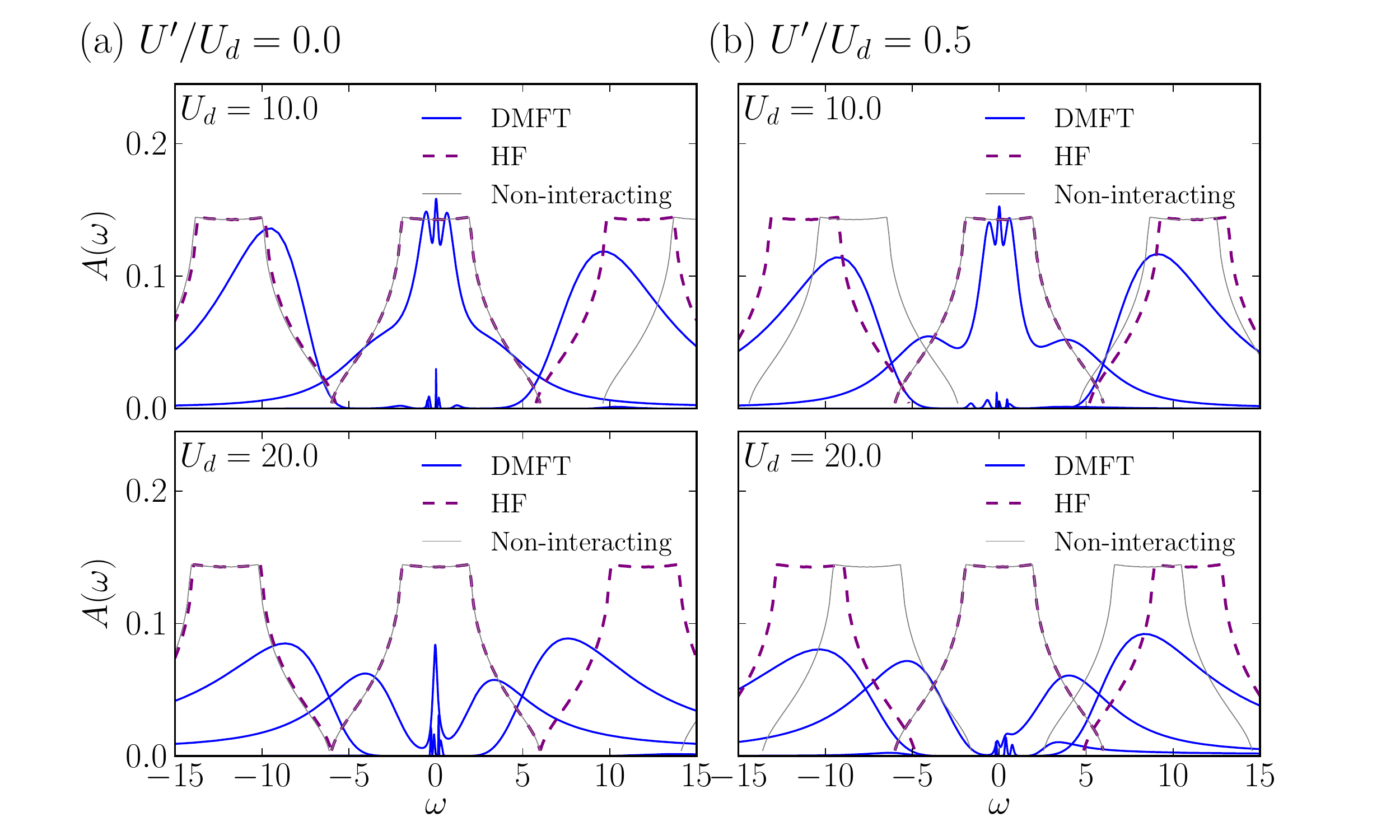}
 \caption{
 (Color online)
 Spectral functions projected on the band basis.
 The data were obtained by solving the three-orbital model with $\Delta=10$ and $t^\prime=4$ at $\beta=15$.
 The solid thick lines denote the spectrum function obtained by analytical continuation of DMFT data.
 The Hartree-Fock and non-interacting band structures are shown by broken lines and thin gray lines, respectively.
 }
 \label{fig:3orb-gap10-spectrum}
\end{figure}

First, we investigate the metal-insulator transition by changing $\Ud$ for fixed $U'/U_d$.
We compare the quasi-particle weights of the four models in Fig.~\ref{fig:3orb-gap10-z}.
Let us first look at the results for $U^\prime/U_d=0.0$ in Fig.~\ref{fig:3orb-gap10-z}(a).
The full model exhibits a metal-insulator transition at $U_d\simeq 20$.
In the metallic phase, we see that the quasi-particle weight of the full model and the unscreened model are almost identical, which indicates small screening effects.
The spin-cRPA model well reproduces the quasi-particle weights in the metallic phase as well as the critical value of the transition.
On the other hand, the quasi-particle weights are substantially overestimated by the charge-cRPA model in the metallic phase.
Furthermore, the critical value of the transition is overestimated by about 25 \% by the charge-cRPA model. 

As shown in Figs.~\ref{fig:3orb-gap10-z}(b) and~\ref{fig:3orb-gap10-z}(c), in the metallic phase, 
the quasi-particle weights of the full model become considerably larger than those of the unscreened model.
This clearly illustrates the enhancement of screening effects by $U^\prime$.
This trend is not reproduced by the spin-cRPA model.
In other words, the screening effects in the spin-cRPA stays negligibly small in the figure.
For $U^\prime/U_d$=0.25, the charge-cRPA method gives a better agreement with the full model compared to the spin-cRPA method.
This agreement is just accidental because the quasi-particle weights of the charge-cRPA model overshoot those of the full model as $U^\prime/U_d$ increases:
The charge-cRPA method underestimates the quasi-particle weights for $U^\prime/U_d=0.5$.
For $U^\prime/U_d=0.5$, we also show the results obtained by the spin-cRPA static model.
This static model reproduces the results of the spin-cRPA model with the dynamical $U$ even near the Mott transition.

We see a similar trend when looking at the Green's function on the Matsubara axis (Fig.~\ref{fig:3orb-gap10-g-comp}).
For $U^\prime/U_d=0$, the data obtained by all the models almost fall on the same curve at $U_d=0$, where the screening effects are small.
However, as $U_d$ increases, the data for the charge-cRPA model shows a more metallic behavior compared to the full model.
This is consistent with the trend in the quasi-particle weights in Fig.~\ref{fig:3orb-gap10-z}(a).
For $U^\prime/U_d=0.5$, the Green's function for the full model is substantially more metallic compared to the unscreened model.
However, this is captured neither by the spin-cRPA nor by the charge-cRPA low-energy models.

Now next look at how the violation of the Pauli principle leads to the overscreening effects for small $U^\prime/U_d$.
For this, we consider the dispersionless two-orbital model illustrated in Fig.~\ref{fig:two-orbital-model}(a).
The Hamiltonian reads
\begin{eqnarray}
  \mathcal{H} &=& \sum_\alpha E_\alpha \hat{n}_{\alpha}-t^\prime \sum_\sigma \left(\hat{c}^\dagger_{1\sigma}\hat{c}_{2\sigma} + \hat{c}^\dagger_{2\sigma}\hat{c}_{1\sigma} \right)\nonumber\\
  &&+\sum_{\alpha} U_\alpha \hat{n}_{\alpha\uparrow}\hat{n}_{\alpha\downarrow}.\label{eq:two-orb-Ham}
\end{eqnarray}
We assume that the system contains one electron (canonical ensemble).
We show the first few diagrams in the charge-cRPA expansion for the on-site interaction on the target band schematically in Fig.~\ref{fig:two-orbital-model}(b).
Considering that $U_d$ acts between different spins and each interaction line flips the spin,
the expansion must contain only odd-order diagrams, e.g., $O(U_d^2)$ and $O(U_d^4)$.
The unphysical first diagram gives the following contribution to the screened interaction on the target band:
\begin{eqnarray}
 U(\omega=0) - U(\omega=\infty) &=& \left\{-\frac{ 2{t^\prime}^2}{\Delta^3}+O({t^\prime}^3) \right\}U^2 \nonumber\\
 && + O({t^\prime}^4U^3),
\end{eqnarray}
which amounts to a screening effect. 
This is essentially the origin of the overscreening seen for the three-orbital model.

Next, we have a look at the spectral function to see if the Hartree-Fock basis is an appropriate choice.
The spectral function is computed by using the maximum-entropy analytic continuation method.~\cite{Jarrell:1996uo}
As seen in the data for $U_d=10$, the Coulomb interactions substantially change the relative position of the screening bands from those of the non-interacting band structure.
Note that the non-interacting band structure is not particle-hole asymmetric for $U_d>0$ due to $E^\mathrm{dc}_\alpha$ in Eq.~(\ref{eq:Ham}).
The positions of the screening bands are however well reproduced by the Hartree-Fock calculations in the metallic phase, i.e., $U_d=10$.
The agreement becomes worse as we get close to the Mott transition for $U^\prime/U_d=0$ where the band shifts by the Coulomb interaction are large.
This indicates that the Hartree-Fock basis might not be an appropriate basis near the Mott transition.
\begin{figure}[b]
 \centering\includegraphics[width=.48\textwidth,clip]{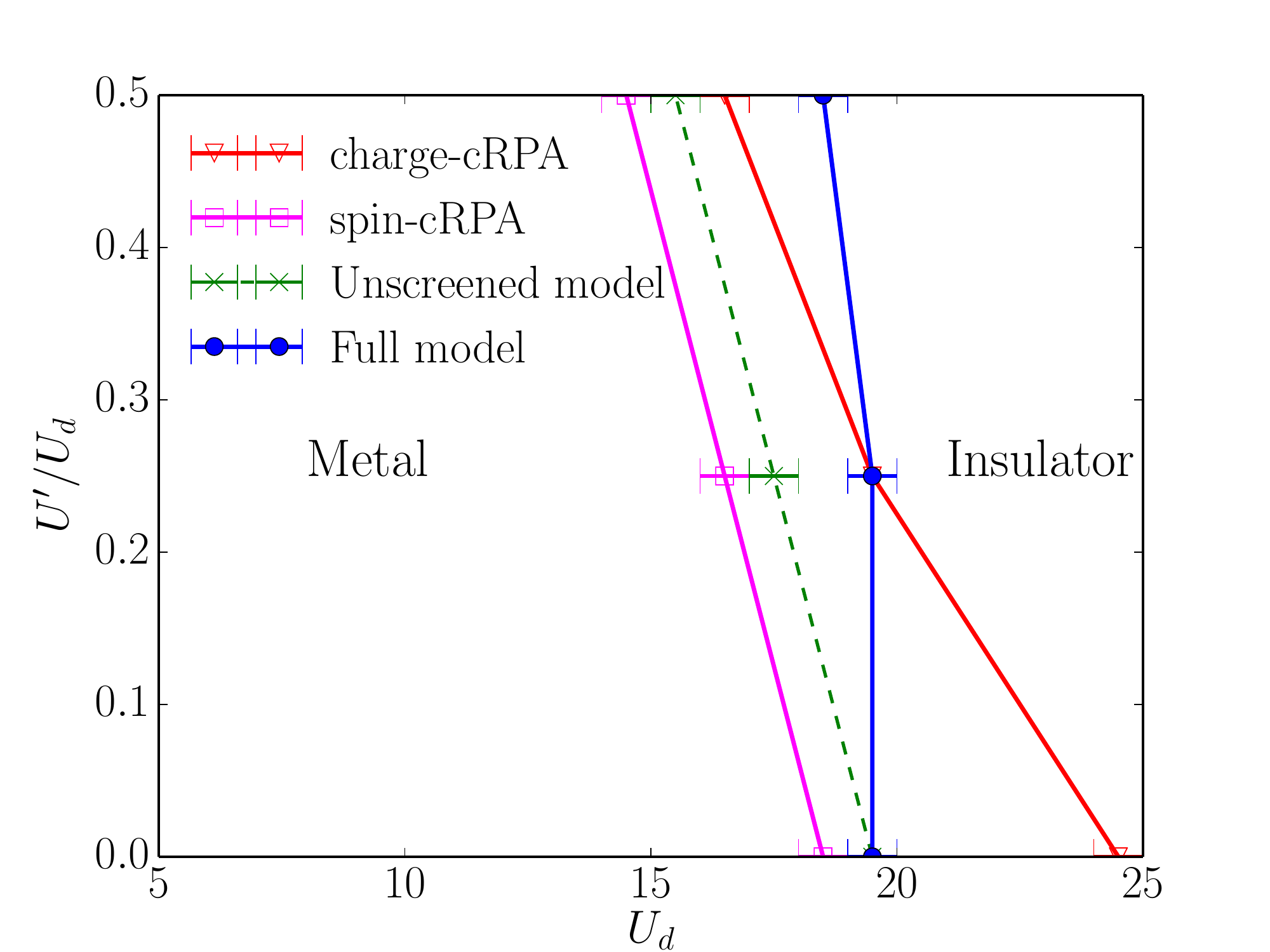}
 \caption{
 (Color online)
 Phase diagram of the three-orbital model for $\Delta=10$ and $t^\prime=4$ at $\beta=15$.
 The solid lines denote the Mott-transition lines for the charge-cRPA (triangle), spin-cRPA (square), unscreened (cross), and full models (filled circle). 
 }
 \label{fig:3orb-gap10-pd}
\end{figure}

We summarize our results in a phase diagram shown in Fig.~\ref{fig:3orb-gap10-pd}.
The Mott transition is identified by the vanishing of the quasi-particle weight.
The critical value of the Mott transition is overestimated by the charge-cRPA model for $U^\prime/U_d=0$.
Although we see a rather good agreement between the cRPA model and the full mode at $U^\prime/U_d=0.5$, this may be only accidental because the $U^\prime/U_d$ dependence is not correctly capture by the charge-cRPA model.
On the other hand, the spin-cRPA model successfully removes the overscreening effects by the violation of the Pauli principle at $U^\prime/U_d=0$.
However, the $U^\prime/U_d$ dependence is not correctly reproduced.

\section{Discussion and Conclusion}\label{sec:summary}
We compared the low-energy properties for the three-orbital model and the corresponding downfolded models obtained by  two variants of the cRPA method (charge-cRPA and spin-cRPA).
The screened Coulomb interactions were projected onto the target band near the Fermi level in a Hartree-Fock band structure. 
We have found that the charge-cRPA method shows overscreening in the parameter region where the intra-orbital repulsion $U^\prime$ is small.
Analyzing a simplified dispersionless two-orbital model,
the origin can be ascribed to the violation of the Pauli principle in the diagrammatic expansion.
We have shown that the spin-cRPA method successfully removes this overscreening.
However, the spin-cRPA method does not correctly reproduce the $U^\prime$-dependence of the Mott transition point for the full three-orbital model.
In particular, the spin-cRPA method show a small antiscreening effect, 
while the full model exhibits substantial screening effects when $U^\prime/U_d$ is large.
We furthermore found a good agreement between the positions of the screening bands in the DMFT spectral function and those obtained by the Hartree-Fock approximation which was used for constructing the Wannier function.
However, this agreement becomes worse near the Mott insulator.

Let us briefly discuss possible origins of the disagreement between the full model and the spin-cRPA model for large $U^\prime/U_d$.
First, the RPA diagrams are not generally the most dominant ones at each expansion order~\cite{Honerkamp:2012bm,Kinza:2015uh} since our model contains only short-ranged interactions.
Thus, the RPA method could miss diagrams which substantially contribute to the screening.
Another issue is the choice of the target manifold.
In the present study, we computed the polarization function
and constructed the Wannier functions based on the Hartree-Fock band structure.
This mean-field basis might not be accurate enough, especially near the Mott transition. 
In the present study, we ignore the renormalization of the kinetic energy by the downfolding, as is done in first-principles calculations.
More elaborated scheme such as the GW method~\cite{Hedin:1965tu} could capture at least some of the correlation-induced shifts and renormalizations of the target and screening bands.

Before closing this paper, we discuss possible future studies.
For the present three-orbital model, we observed antiscreening effects in the parameter regime considered.
To realize a large screening effect, we may have to increase the number of screening bands. A five-orbital set-up has already been considered but these results were similar to those shown here for the three-orbital case.~\cite{downfoldingv1} 
Treating a substantially larger number of screening bands may not be feasible for three-dimensional models, 
because the computational complexity of solving the quantum impurity problem scales exponentially in the number of orbitals.
A possible future direction is testing the downfolding scheme for one dimensional problems, where a full model with many screening bands could be solved exactly by lattice quantum Monte Carlo.
In this set-up, one may also be able to examine the role of long-range Coulomb interactions.

A recently proposed generalization of the cRPA scheme is 
the Wick-ordered constrained functional renormalization group (cfRG) method.~\cite{Honerkamp:2012bm,Kinza:2015uh}
This scheme has been tested for one- and two-dimensional models with a few screening bands and one target band,~\cite{Honerkamp:2012bm,Kinza:2015uh}
which is similar to our set-up. 
The cfRG calculations revealed relevant and qualitative corrections to the effective interactions beyond cRPA.
More extensive test of this method will be interesting.
It will furthermore be interesting to examine to what extent the violation of the Pauli principle affects the screened interactions computed for real compounds.
For example, in the case of high-$T_c$ cuprates, there are $p$ bands close to the Fermi level,
a situation which resembles the configuration of the few-orbital model with narrow gaps considered in this study.

\begin{acknowledgments}
We thank Rei Sakuma for explanations on the product basis and
Fakher Assaad, Ferdi Aryasetiawan, Shintaro Hoshino, Masatoshi Imada, Takashi Miyake, Kazuma Nakamura and Shiro Sakai for stimulating discussions and useful comments. 
We acknowledge support from the DFG via FOR 1346, the SNF Grant 200021E-149122, ERC Advanced Grant SIMCOFE and NCCR MARVEL.
The calculations have been performed on the M\"{o}nch and Brutus clusters of ETH Z\"{u}rich using codes based on ALPS.~\cite{Bauer:2011tz}
\end{acknowledgments}

\bibliography{ref,ref2}
\appendix

\section{Orbital-dependent mean fields}\label{sec:band-shift}
\begin{figure}
 \centering\includegraphics[width=.4\textwidth,clip,type=pdf,ext=.pdf,read=.pdf]{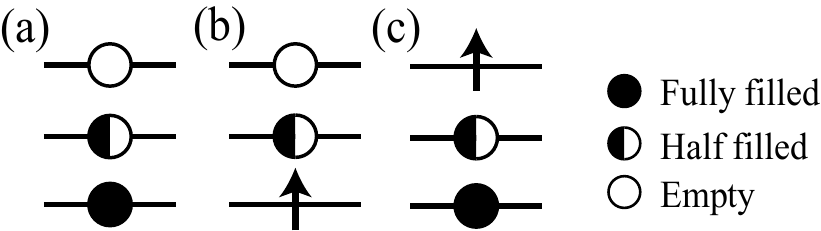}
 \caption{Single-particle excitations in the atomic limit: (a) the ground state, (b) the excited state with an additional hole in the lowest orbital, and (c) the excited state with an additional spin in the highest orbital.
 We assume that half-filled target orbitals are paramagnetic.
 }
 \label{fig:band-shift}
\end{figure}

Our model given in Eq.~(\ref{eq:Ham}) breaks the particle-hole symmetry for nonzero interaction. 
As a consequence, orbital-dependent mean fields change the relative position of the screening bands.
The orbital-depedent chemical potentials $E^\mathrm{dc}_\alpha$ in Eq.~(\ref{eq:Ham}) is introduced to cancel this band shift  in the atomic limit, i.e., for $t^\prime=0$ and $t=0$.
In the ground state, the three orbitals are filled, half filled, and empty, respectively [see illustration in Fig.~\ref{fig:band-shift}(a)].
First, we remove a spin from the lowest orbital [see Fig.~\ref{fig:band-shift}(b)].
This excitation costs
\begin{eqnarray}
  \Delta E_\mathrm{hole} &=& -U_r-U^\prime+\mu +\Delta = U^\prime +\Delta,
\end{eqnarray}
where $\mu$ ($=U_d/2+2U^\prime$) is the chemical potential and $U_r~(=U_d/2)$ is the on-site Coulomb interaction on the screening orbitals.
On the other hand, putting an electron 
into the highest orbital results in the excited state shown in Fig.~\ref{fig:band-shift}(c).
The excitation energy is given by
\begin{eqnarray}
  \Delta E_\mathrm{electron} &=& 3U^\prime - \mu + \Delta = U^\prime - \frac{U_d}{2} + \Delta.\label{eq:Eelec}
\end{eqnarray}
$\Delta E_\mathrm{hole}$ and $\Delta E_\mathrm{electron}$ correspond to the positions of the lower and upper screening bands in the spectral function, respectively.
When we increase $U_d$ with $U^\prime/U_d$ fixed, $\Delta E_\mathrm{hole}$ increases linearly with $U_d$.
On the electron side, $\Delta E_\mathrm{electron}$ stays constant for $U^\prime=U_d/2$ or decreases for $U^\prime<U_d/2$ as $U_d$ increases.
To cancel out this band shift, we take 
$E^\mathrm{dc}_\alpha=U^\prime, 0, -U^\prime+U_d/2$ for $\alpha=1,2,3$.

\section{Rotating the single-particle basis when solving a multi-orbital impurity problem}\label{appendix:basis-rot}
We consider a multi-orbital quantum impurity problem given by the action 
\begin{eqnarray}
  S &=& S_\mathrm{imp} + \sum_{ab} \int_0^\beta  \diff \tau \diff \tau^\prime \Delta_{ab}(\tau-\tau^\prime) c^\dagger_a(\tau) c_b(\tau^\prime),\hspace{5mm}\label{eq:action-cthyb}
\end{eqnarray}
where $\Delta$ is the hybridization function, which satisfies $\Delta_{ab}(\tau) = \Delta_{ba}^*(\tau)$.
$S_\mathrm{imp}$ is the local impurity action. 

One expands the partition function $Z$ as \begin{eqnarray}
  Z &=& \mathrm{Tr} \left[ e^{-\beta \mathcal{H}}\right]\nonumber \\
  &=& Z_\mathrm{bath} \sum_{n=0}^\infty \int_0^\beta \mathrm{d} \tau_1 \mathrm{d} \tau_1^\prime \cdots \int_0^\beta \mathrm{d} \tau_n \mathrm{d} \tau_n^\prime \nonumber\\
   && \mathrm{Tr_{loc}}\left[ e^{-\beta\mathcal{H}_\mathrm{loc}} T \opc_{\alpha_n}(\tau_n) \opcdag_{\alpha_n^\prime}(\tau_n^\prime) \cdots \opc_{\alpha_1}(\tau_1) \opcdag_{\alpha_1^\prime}(\tau_1^\prime)\right]\nonumber\\
   && \times \mathrm{det} \boldsymbol{M}^{-1},\label{eq:z-cthyb}
\end{eqnarray}
where $H$ is the Hamiltonian of the whole system including the impurity and the bath.
$\mathcal{H}_\mathrm{loc}$ is the local Hamiltonian corresponding to $S_\mathrm{imp}$.
The matrix element of $\boldsymbol{M}^{-1}$ at $(i,j)$ is given by the hybridization function $\Delta_{\alpha_i^\prime,\alpha_j}(\tau_i^\prime - \tau_j)$.
In the Krylov method, we evaluate the trace over the local degrees freedom $\mathrm{Tr}_\mathrm{loc}[\cdots]$ by calculating imaginary time evolutions in the occupation number basis.~\cite{Lauchli:2009er}

When $\Delta$ has non-vanishing off-diagonal elements $\Delta_{ab}$ ($a\neq b$), the 
Monte Carlo sampling according to Eq.~(\ref{eq:z-cthyb}) suffers from a negative sign problem.
To reduce this sign problem, we rewrite the action Eq.~(\ref{eq:action-cthyb}) as
\begin{eqnarray}
  S &=& S_\mathrm{imp} + \sum_{ab} \int_0^\beta  \diff \tau \diff \tau^\prime \bar{\Delta}(\tau-\tau^\prime) d^\dagger_a(\tau) d_b(\tau^\prime),\hspace{3mm}\label{eq:action-cthyb-rot}
\end{eqnarray}
where
\begin{eqnarray}
  c_a(\tau) &=& \sum_b U_{ab} d_b(\tau),\\
  c^\dagger_a(\tau) &=& \sum_b (U^\dagger)_{ab} d^\dagger_b(\tau),\\
  \bar{\Delta}_{ab} (\tau) &=& \sum_{cd} (U^\dagger)_{ac} \Delta_{cd}(\tau-\tau^\prime) U_{db},
\end{eqnarray}
and $U_{ab}$ is a unitary matrix.
We choose the unitary matrix $U$ such that the off-diagonal elements of $\bar{\Delta}$ become smaller.
In the present study, we choose the single-particle basis that diagonalizes the non-interacting part of $\mathcal{H}_\mathrm{imp}$
because this diagonalizes the hybridization function at all $\tau$ 
in the non-interacting limit, i.e, $U_d=U^\prime=0$.

The partition function is then expanded in terms of this new basis as
\begin{eqnarray}
  Z &=& Z_\mathrm{bath} \sum_{n=0}^\infty \int_0^\beta \mathrm{d} \tau_1 \mathrm{d} \tau_1^\prime \cdots \int_0^\beta \mathrm{d} \tau_n \mathrm{d} \tau_n^\prime \nonumber\\
   && \mathrm{Tr_{loc}}\left[ e^{-\beta\mathcal{H}_\mathrm{loc}} T d_{\alpha_n}(\tau_n) d^\dagger_{\alpha_n^\prime}(\tau_n^\prime) \cdots d_{\alpha_1}(\tau_1) d^\dagger_{\alpha_1^\prime}(\tau_1^\prime)\right]\nonumber\\
   && \times \mathrm{det} \boldsymbol{\bar{M}}^{-1},\label{eq:z-cthyb2}
\end{eqnarray}
where the matrix element of $\boldsymbol{\bar{M}}^{-1}$ is now given by the rotated hybridization function $\bar{\Delta}$.
The local trace can be efficiently evaluated in the occupation number basis.

\end{document}